\documentclass[twocolumn,pra,twocolumn,superscriptaddress]{revtex4}
\usepackage[latin9]{inputenc}
\usepackage{color}
\usepackage{amsmath}
\usepackage{amssymb}\usepackage{amsmath}
\usepackage{amsfonts}
\usepackage{amssymb}
\usepackage{graphicx}
\usepackage{tikz}
\usepackage{mathrsfs} % cal F fidelity
\usepackage{float}
\usepackage{bbm}
\usepackage{epstopdf}
\usepackage{epsfig}
\usepackage{verbatim}
\usepackage{array}
\usepackage{setspace}
\usepackage{times}
\usepackage{tabularx}
\usepackage{setspace}
\usepackage{ulem}
\usepackage{soul}
\usepackage[unicode=true,
bookmarks=true,bookmarksnumbered=false,bookmarksopen=false,
breaklinks=false,pdfborder={0 0 1},backref=false,colorlinks=true]
{hyperref}
\hypersetup{
	linkcolor=magenta, urlcolor=blue, citecolor=blue, pdfstartview={FitH}, hyperfootnotes=false, unicode=true}

\newcolumntype{C}[1]{>{\centering\arraybackslash$}p{#1}<{$}}

\begin{document}

\title{Fast high-fidelity geometric gates for singlet-triplet qubits}

\author{Mei-Ya Chen}
\affiliation{Guangdong Provincial Key Laboratory of Quantum Engineering and Quantum Materials,
	and School of Physics\\ and Telecommunication Engineering, South China Normal University, Guangzhou  510006, China}

\author{Chengxian Zhang}\email{cxzhang@gxu.edu.cn}
\affiliation{ School of Physical Science and Technology, Guangxi University, Nanning 530004, China}

\author{Zheng-Yuan Xue}  \email{zyxue83@163.com}
\affiliation{Guangdong Provincial Key Laboratory of Quantum Engineering and Quantum Materials,
	and School of Physics\\ and Telecommunication Engineering, South China Normal University, Guangzhou  510006, China}
\affiliation{Guangdong-Hong Kong Joint Laboratory of Quantum Matter and  Frontier Research Institute for Physics,\\ South China Normal University, Guangzhou  510006, China}

\date{\today}

\begin{abstract}
	
Geometric gates that use the global property of the geometric phase is believed to be a powerful tool to  realize fault-tolerant quantum computation. However, for singlet-triplet qubits in semiconductor quantum dot, the low Rabi frequency of the microwave control leads to overly long gating time, and thus the constructing geometric gate suffers more from the decoherence effect. Here we investigate the key issue of whether the fast geometric gate can be realized for singlet-triplet qubits without introducing an extra microwave-driven pulse, while maintaining the high-fidelity gate operation at the same time. We surprisingly find that both the single- and two-qubit geometric gates can be implemented via only modulating the time-dependent exchange interaction of the Hamiltonian, which can typically be on the order of $\sim$GHz, and thus the corresponding gate time is of several nanoseconds. Furthermore, the obtained geometric gates are superior to their counterparts, i.e., the conventional dynamical gates for singlet-triplet qubits, with a relatively high fidelity surpassing 99\%. Therefore our scheme is particularly applied to singlet-triplet qubits to obtain fast and high-fidelity geometric gates. Our scheme can also be extended to other systems without a microwave drive.

\end{abstract}

\maketitle

\section{Introduction}

Spin qubits in semiconductor quantum dots are a promising technique to realize universal quantum computation due to their potential scalability via combining the quantum technologies with the state-of-the-art semiconductor industry \cite{Chatterjee.2021}. Currently, fast and high-fidelity control is still crucial for spin qubits to realize fault-tolerant quantum tasks. Thus various types of qubits using the electron spin degree of freedom in quantum dot systems have been tested over recent years, including single-dot spin qubits \cite{Loss.1998,Veldhorst.2014, Huang.2019,Noiri.2022}, double-dot singlet-triplet  qubits \cite{Petta.2005,Foletti.2009, Zhang.2017, Abadillo.2019, Pascal.2020,Federico.2021}, as well as hybrid qubits \cite{Shi.2012, Frees.2019}, triple-dot exchange-only qubits \cite{Divincenzo.2000, Laird.2010}, and resonant qubits \cite{Taylor.2013,Medford.2013b}.

Among these possible candidates, singlet-triplet qubits are particularly standing out, since their use can implement all-electrical gate operation control with long coherence times and fast gate operation. Gate operation for singlet-triplet qubits is implemented via tuning the Heisenberg exchange interaction between the two spins with gate times on the order of nanoseconds, thanks to the large exchange interaction ($\sim$ GHz) \cite{Petta.2005}. Recently, researchers have experimentally observed that the relaxation time of the spin states in a quantum dot can be as long as 9 s, and the Overhauser noise can be substantially suppressed \cite{Ciriano-Tejel.2021} by using the isotopic purification technique in the silicon platform. Nevertheless, singlet-triplet qubits in a quantum dot setup are still sensitive to the background charge noise, which occurs in the vicinity of the quantum dot \cite{Bermeister.14,Chan.18}. This hinders high-fidelity operation for either single or two-qubit quantum gates. Much work has been devoted to mitigating the charge noise, including working near the charge noise sweet spots \cite{Martins.2016} and designing gates using dynamical corrected gates \cite{Wang.2014,Throckmorton.17}.

Alternatively, the geometric phase  \cite{Berry.84} is believed to be useful to combat the noise effect. After a cyclic evolution in the parameter space, the quantum state can acquire an extra global phase factor, i.e., the Berry phase, under the adiabatic condition. Inspired by Berry's idea, it is realized that the global property of the geometric phase can be a powerful tool for quantum computation \cite{Pachos.99,Zanardi.99,Duan.01,Zhu.02,Zhu.03}. 
%Since then, geometric quantum gates based on adiabatic evolution have attracted much attentions.
However, the adiabatic condition hinders  wide application of  geometric gates owing to the overly long evolution time, which renders more decoherence. Recently, a universal set of quantum gates based on the nonadiabatic geometric phase, namely, the Aharonov-Anandan phase \cite{Aharonov.87}, has been realized in superconducting qubits \cite{abdumalikov.13, xu.18,Tao.18, Liu.19, egger.19, xu.20,Xujing.20,Tao.20,Lisai20,Ding.21a,Ding.21b}, trapped ions \cite{ai.20,ai.21,Guo.2021}, semiconductor quantum dots \cite{Solinas.03a, Mousolou.14, Mousolou.17b,Zhang.2020}, etc. The point of constructing a geometric gate is to cancel out the accompanied dynamical phase during the cyclic evolution, leaving only the wanted geometric phase. Typically, this can be realized by introducing a microwave field to operate its time-dependent phase to ensure that the quantum state is always evolving along the longitude of the Bloch sphere \cite{Zhao.17,Zhang.2020}.

By applying microwave-driven pulses on the detuning value, a recent experiment \cite{Takeda.20} showed 99.6\% single-qubit gate fidelity for a singlet-triplet qubit in a silicon-based semiconductor quantum dot. On the other hand, the Rabi frequency reported there is with only several MHz. This small value for the singlet-triplet system is also comparable to the typical values using the electric-dipole spin resonance technology in similar devices for single-dot spin qubits \cite{Yoneda.18}.%On the other hand,  the Rabi frequency of a quantum system controlled by microwave   is limited in experiments. For example,  in silicon-based semiconductor quantum dot for \blue{singlet-triplet} qubit, the exchange Rabi frequency is only several MHz \cite{Takeda.20}, which is comparable to the typical values using the EDSR technology in similar devices for single-dot spin qubits \cite{Yoneda.18}. 
Therefore the gate time of the desired geometric gate can be typically on the order of microseconds, which is much longer than the traditional dynamical gate without using the microwave field. To fully employ the advantage of  geometric gates, one has to seek ways to enable fast and appropriate operation. A good compromise is to use the Landau-Zener interferometry \cite{Shevchenko.10,Stehlik.12,Wang.16} with respect to the quantum state to fulfill cyclic evolution using the dc-gating pulse. However, a recent experiment \cite{Wang.16} indicates that the  dynamical phase is difficult to remove and a more complicated technique like spin echo is needed, making it impractical for quantum computation.

Here we propose a framework to realize both single- and two-qubit nonadiabatic geometric gates without the external microwave field so that the gate time can be only several nanoseconds. By only modulating the time-dependent exchange interaction, the quantum state in the parameter space can evolve along the specific geodesic line where no dynamical phase will be introduced. Thus our method is simple but experimentally feasible. The avoidance of using a microwave field not only enables short gate duration but can also simplify the control complexity for the system. This could be another potential advantage for this approach, especially considering power dissipation and addressability in the context of scaling to large qubit numbers \cite{Pascal.2020}. By numerically performing randomized benchmarking \cite{Emerson.2005,Knill.2008,Easwar.2012} and calculating the filter function \cite{Green.2012,Green.2013,Paz-Silva.2014} under the realistic $1/f$ charge noise environment, we surprisingly find that all gate fidelities can be higher than 99\%, which surpasses the conventional dynamical gate. Our results indicate that singlet-triplet qubits might benefit from the preservation of the geometric operation to obtain high-fidelity control.

We emphasize that our method is not only suited to the exchange-coupled singlet-triplet spin qubits but also can be readily extended to other systems that can be described by the Ising-type interacting Hamiltonian \cite{Buterakos.21}, such as the superconducting transmon qubits \cite{Collodo.20} and the capacitively coupled charge qubits \cite{Shinkai.09}.

\begin{figure}[bp]
	\includegraphics[width=0.9\columnwidth]{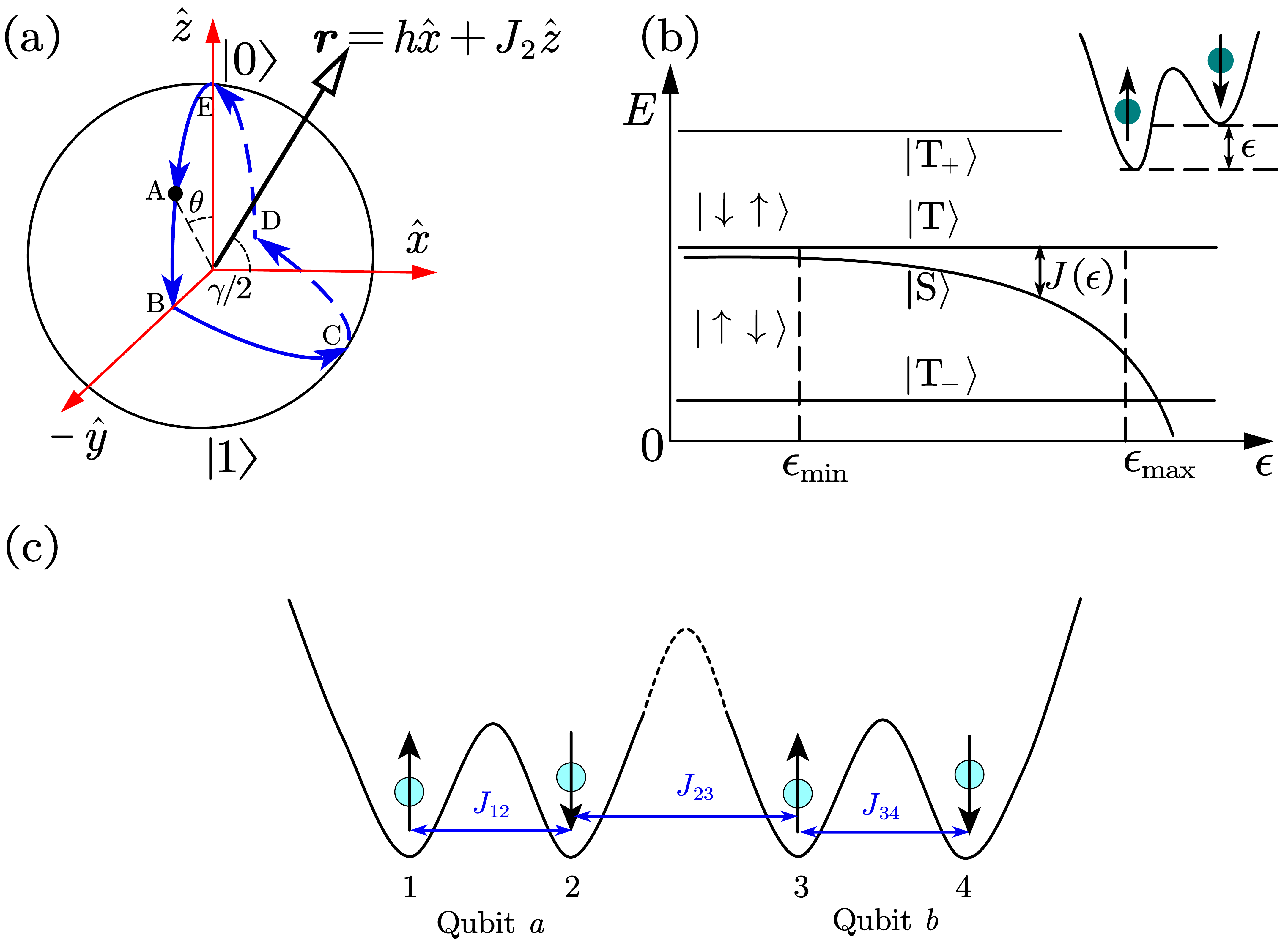}
	\caption{(a) The schematic of the evolution path to induce a geometric gate. The dressed state $|\psi_{+}\rangle$ evolves along the cyclic path A-B-C-D-E-A to obtain the global geometric phase so as to get a desired geometric gate. The normal vector (also, the rotation axis) with respect to the plane B-C-D is denoted as $\boldsymbol{r}=h \hat{x}+J_{2} \hat{z}$. The angle between the vector and the $x$ axis is $\gamma/2$. (b) Energy level of the double-dot system as a function of the detuning $\epsilon$, used to control the exchange interaction $J$. (c) A lateral four-quantum-dot system with each dot labeled by 1, 2, 3, and 4, from the left to the right to enable two-qubit operation for singlet-triplet qubits, where dots 1 and 2 form qubit a, and dots 3 and 4 form qubit b. The quantum dots are coupled via the exchange interaction denoted by $J_{i,i+1}$ ($i=1,2,3$).}	
	\label{fig:path}
\end{figure}

\section{MODEL}\label{sec:model}

The control Hamiltonian for a singlet-triplet qubit is \cite{Wang.2014}
\begin{equation}
  H_{\rm{ST}}(t)=\frac{h}{2}\sigma_x+\frac{J[\epsilon{(t)}]}{2}\sigma_z,
  \label{eq:Hamiltonian1}
\end{equation}
where $\sigma_x$ and $\sigma_z$ are Pauli matrices. The computational basic states are the spin triplet state $|0\rangle = |\rm{T}(1,1)\rangle = (\left|\uparrow\downarrow\right\rangle+\left|\downarrow \uparrow\right\rangle) / \sqrt{2}$ and the singlet state $ |1\rangle = |\rm{S}(1,1)\rangle = (\left|\uparrow\downarrow\right\rangle-\left|\downarrow\uparrow\right\rangle)/\sqrt{2} $. Here, we define the spin state $\left|\downarrow\uparrow\right\rangle = c_{1 \downarrow}^{\dagger} c_{2\uparrow}^{\dagger}|\mathcal{V}\rangle$, where $c_{i \tau}^{\dagger}$ ($i=1,2$) denotes creating an electron with spin $\tau$ at the $i$th quantum dot, and $|\mathcal{V}\rangle$ denotes the vacuum state. $h=g \mu \Delta B$ refers to the magnetic field gradient across the two quantum dots, where $g$ is the electron $g$ factor, $\mu$ is the Bohr magneton, and $\Delta B$ denotes the difference of the magnetic field between the double quantum dots. Experimentally, $h$ can set to be any desired constant value from several MHz to $\sim$ GHz, by either the dynamical nuclear polarization \cite{Bluhm.10} or micromagnet \cite{Watson.2018} technique, in both GaAs \cite{Nichol.17} and silicon \cite{Watson.2018} heterostructure. The exchange interaction $J[\epsilon{(t)}]$ can be controlled via operating the detuning $\epsilon$ with respect to the gate voltage, which refers to the energy splitting between the spin singlet state $ |\rm{S}(1,1)\rangle$ and the triplet state $ |\rm{T}(1,1)\rangle$, as seen in Fig.~\ref{fig:path} (b). Here we consider a phenomenological model  $J[\epsilon{(t)}]=J_{0}\mathrm{exp}[\epsilon(t)/\epsilon_{0}]$, which is well fitted from the experimental data for both GaAs \cite{Dial.2013,Cerfontaine.14,Pascal.2020} and silicon \cite{Wu.14} system. According to Ref.~\cite{Cerfontaine.14}, $J_{0}=1\ \mathrm{ns}^{-1}$, $\epsilon_{0}=0.272\ \mathrm{meV}$, and the detuning is constrained as $-5\epsilon_{0}<\epsilon<5\epsilon_{0}$, which implies $0.007J_{0}\leqslant J \leqslant 148 J_{0}$ . In our simulation we consider $J_{\mathrm{min}}\ll h$, such that we assume $0\leqslant J \leqslant J_{\rm{max}}$. Since both $J$ and $h$ can be obtained to be on the order of $\sim$ GHz, the time induced from such coupling is with the order of nanosecond.

In the absence of noise, the Hamiltonian in Eq. (\ref{eq:Hamiltonian1}) leads to the rotation with the form as
\begin{equation}\label{eq:dynamical naive1}
R(J,\phi)=e^{-\frac{i}{\hbar}\frac{h\sigma_x+J\sigma_z}{2}\frac{\phi}{\sqrt{J^2+h^2}}},
\end{equation}
where $R(J,\phi)$ denotes the rotation in the $x$-$z$ plane by an angle $\phi$, and the rotation axis in the plane is determined by $J/h$. In this work we assume $J$ as a square pulse, namely, $H_{\rm{ST}}(t)$ is a piecewise Hamiltonian and therefore $R(J,\phi)$ is a one-piece rotation in a specific time duration. Hereafter, we define $U(\boldsymbol{r}, \phi)=\exp \left(-i \frac{\boldsymbol{\sigma} \cdot \boldsymbol{r}}{|\boldsymbol{r}|} \frac{\phi}{2}\right)$ as rotation around an axis defined by vector $\boldsymbol{r}$. In this way we have $R(J, \phi)=U(h \hat{x}+J \hat{z}, \phi)$.  The rotation out of the $x$-$z$ plane can be implemented by a $z$ rotation sandwiched between two $x$ rotations \cite{nielsen.2002,Zhang.2017}:
\begin{equation}
U(\hat{r},\xi)=U(\hat{x},\xi_1)U(\hat{z},\xi_2)U(\hat{x},\xi_3).
\label{eq:rotation-p1}
\end{equation}
As $h>0$ always exists,  $R(J,\phi)$ cannot implement a single $z$-axis rotation. But a $z$-axis rotation with arbitrary rotation angle can be decomposed into three pieces of composite sequences (the Hadamard-$x$-Hadamard sequence) \cite{Wang.2014}, i.e.,
\begin{equation}
U(\hat{z},\xi_0)=-U(\hat{x}+\hat{z},\pi)U(\hat{x},\xi_0)U(\hat{x}+\hat{z},\pi).
\label{eq:rotation-p2}
\end{equation}
Then, by inserting Eq.~(\ref{eq:rotation-p2}) into Eq.~(\ref{eq:rotation-p1}), one finds
\begin{small}
\begin{equation}
U(\hat{r},\xi)=U(\hat{x},\xi_1)U(\hat{x}+\hat{z},\pi)U(\hat{x},\xi_2)U(\hat{x}+\hat{z}, \pi)U(\hat{x},\xi_3).
\label{eq:dynamical naive}
\end{equation}
\end{small}

The rotation implemented by the pulse  in Eq.~(\ref{eq:Hamiltonian1}) is sensitive to noise. There are mainly two noise sources resulting in gate error. One is the charge noise, which brings fluctuation to the detuning ($\delta \epsilon$) and further leads to the error in the exchange interaction labeled by $\delta J =g[J]\delta\epsilon$, with $g[J] \propto J$ \cite{Wang.2014}. Another is the Overhauser (nuclear spin) noise, which is a time-dependent fluctuation in the background of the nuclear spin bath, adding a small term into the Hamiltonian: $h\rightarrow h+\delta h$. A recent experiment based on the GaAs system indicated that, the standard deviation for Overhauser noise is about $\sigma_{\delta h}=2 \pi \times 2\ \mathrm{MHz}$, when taking $h=\ 2 \pi \times 40\ \mathrm{MHz}$ \cite{Pascal.2020}. In a silicon-based semiconductor quantum dot, the isotopic purification technique can strongly suppress the Overhauser noise \cite{Huang.2019}. A previous work \cite{Kalra.14} has shown that it can be as low as $\delta h / h=2 \times 10^{-5}$. In this work we focus on the silicon system and therefore neglect the Overhauser noise effect. Meanwhile, in a silicon heterostructure it would introduce the unwanted valley-spin coupling leading to relaxation \cite{Klinovaja.2012,Jock.2022}. According to the latest experiment \cite{Ciriano-Tejel.2021}, the relaxation time with the state-of-the-art silicon-based platform has reached $T_1=9 \ \rm{s}$. In this way we also neglect the relaxation effect and assume the evolution is unitary. Considering the noise effect, below any mentioned rotation $U(\hat{r},\xi)$ or $R(J,\phi)$ is with the error term. In the following we call them the naive dynamical gates. In this work our aim is to design the geometric gate to improve the naive dynamical gates. %, as shown later in section.~\ref{sec:results}.

%\section{RESULT}\label{sec:results}
\section{Single-qubit geometric gates} \label{sec:singlegeo}

Here we introduce how to construct the geometric gate via the control Hamiltonian in Eq.~(\ref{eq:Hamiltonian1}), %. To implement the geometric gate, 
and the evolution time can be generally separated into three parts. The Hamiltonian in each interval should satisfy
\begin{equation}
\begin{array}{lll}
H_{1}(t)=\frac{h}{2}\sigma_x+\frac{J_1}{2}\sigma_z,\  \ & T_{\rm{A}} \leqslant t\leqslant T_{\rm{B}} \\
H_{2}(t)=\frac{h}{2}\sigma_x+\frac{J_2}{2}\sigma_z,\  \ & T_{\rm{B}} < t \leqslant T_{\rm{D}} \\
H_{3}(t)=\frac{h}{2}\sigma_x+\frac{J_3}{2}\sigma_z,\  \ & T_{\rm{D}} < t\leqslant T,
\end{array}
\label {eq:Hamiltoniian-paths}
\end{equation}
with $J_{i}({i=1,2,3})$ in each part satisfying
\begin{equation}\label{eq:paths}
\begin{aligned}
&\int_{T_{\rm{A}}}^{T_{\rm{B}}}\sqrt{J_1^2+h^2}dt=\frac{\pi}{2}-\theta+2 n_1 \pi,\ \ & T_{\rm{A}} \leqslant t\leqslant T_{\rm{B}} \\
&\int_{T_{\rm{B}}}^{T_{\rm{D}}}\sqrt{J_2^2+h^2}dt=\pi,\ \ & T_{\rm{B}} < t \leqslant T_{\rm{D}}  \\
&\int_{T_{\rm{D}}}^{T}\sqrt{J_3^2+h^2}dt=\frac{\pi}{2}+\theta+2 n_2 \pi, & T_{\rm{D}} < t\leqslant T.
\end{aligned}
\end{equation}
Here $\theta$ is the parameter determined by the chosen rotation axis, as seen below. $n_i\  (i=1,2)$ depends on the values of $\theta$:
\begin{equation}
n_1=\left\{\begin{array}{ll}
1, & \theta > \pi/2  \\
0, & \theta \leqslant \pi/2  \\
\end{array}\right.\ \ n_2=\left\{\begin{array}{ll}
0, & \theta \geqslant -\pi/2  \\
1, & \theta< -\pi/2  \\
\end{array}\right..
\end{equation}
On the other hand, since $h$ retains a constant value during all the gate operation processing, we take $h=1$ as our energy unit in the remainder of this work. The exchange interaction value is therefore equivalent to the ratio $J/h$.

The corresponding evolution operators in each segment are
\begin{equation}\begin{array}{lll}
	U_1(t,T_{\rm{A}})=e^{-\frac{i}{\hbar} \int_{T_{\rm{A}}}^{t} H_1(t')d t'},\ \ & T_{\rm{A}} \leqslant t\leqslant T_{\rm{B}} \\
	U_2(t,T_{\rm{B}})=e^{-\frac{i}{\hbar} \int_{T_{\rm{B}}}^{t} H_2(t')d t'},\ \ & T_{\rm{B}} < t\leqslant T_{\rm{D}}  \\
    U_3(t,T_{\rm{D}})=e^{-\frac{i}{\hbar} \int_{T_{\rm{D}}}^{t} H_3(t')d t'},\ \ & T_{\rm{D}} < t\leqslant T. \\
	\end{array}
	\label{eq:evo}
\end{equation}
Taking $J_1=J_3$ throughout this work, the total evolution operator at the final time will be
\begin{widetext}
\begin{equation}
	\begin{aligned}
	U_{\rm{g}}(T,{T_{\rm{A}}})
	&={U_3}\left({T},{T_{\rm{D}}}\right){U_2}\left( {T_{\rm{D}}},{T_{\rm{B}}}\right){U_1}\left({T_{\rm{B}}},{T_{\rm{A}}}\right)\\
	&=\left(
	\begin{array}{ccc}
	\frac{i(J_1-J_2)\cos{\theta}-\sqrt{J_1^2+1}(J_1J_2+1)}{(J_1^2+1)\sqrt{J_2^2+1}}   & \frac{(J_2-J_1)(\sqrt{J_1^2+1}\sin{\theta}+iJ_1\cos{\theta})}{(J_1^2+1)\sqrt{J_2^2+1}}   \\
	\frac{(J_1-J_2)(\sqrt{J_1^2+1}\sin{\theta}-iJ_1\cos{\theta})}{(J_1^2+1)\sqrt{J_2^2+1}}   & -\frac{i(J_1-J_2)\cos{\theta}+\sqrt{J_1^2+1}(J_1J_2+1)}{(J_1^2+1)\sqrt{J_2^2+1}}   \\
	\end{array}\right).
	\label{eq:evolution1}
	\end{aligned}
\end{equation}
\end{widetext}
By setting $J_1=0,\ J_2=\tan({\gamma/2})$, we can further simplify $U_{\rm{g}}(T,T_{\rm{A}})$ as
\begin{small}
\begin{equation}\label{eq:evolution2}
\begin{aligned}
U_{\rm{g}}(\hat{r},{\gamma})
&=\left(\begin{array}{ccc}
-\cos{\frac{\gamma}{2}}-i\sin{\frac{\gamma}{2}}\cos{\theta}   & \sin{\frac{\gamma}{2}}\sin{\theta}   \\
-\sin{\frac{\gamma}{2}}\sin{\theta}   & -\cos{\frac{\gamma}{2}}+i\sin{\frac{\gamma}{2}}\cos{\theta}   \\
\end{array}\right)\\
&=-e^{-i \frac{\gamma}{2} ( \sin{\theta} \sigma_y-\cos{\theta} \sigma_z)}.
\end{aligned}
\end{equation}
\end{small}$U_{\rm{g}}({\hat{r},\gamma})$ represents rotation around the axis $\hat{r}=(0,\sin\theta,-\cos\theta)$ on the $y$-$z$ plane by an angle $\gamma$, where the rotation axis is determined by $\theta$. Other rotations that are out of the $y$-$z$ plane can be implemented by using either the sequence $y$-$z$-$y$ or $z$-$y$-$z$, similar to the case in Eq.~(\ref{eq:rotation-p1}).

To demonstrate that $U_{\rm{g}}(\hat{r},{\gamma})$ is the desired geometric gate, we introduce the orthogonal dressed states:
\begin{eqnarray}\label{eq:states}
\left|\psi_{+}\right\rangle&=&\cos\frac{\theta}{2}|0\rangle -i \sin\frac{\theta}{2}|1\rangle, \notag\\
\left|\psi_{-}\right\rangle&=&i\sin\frac{\theta}{2} |0\rangle-\cos\frac{\theta}{2}|1\rangle.
\end{eqnarray}
For an arbitrary operator $U_{\rm{g}}({\hat{r},\gamma})$ with respect to $\theta$, its evolution can be visualized by the Bloch sphere using the dressed states, as shown in Fig.~\ref{fig:path}(a). The evolution of the dressed state $|\psi_{+}\rangle$ (the case for $|\psi_{-}\rangle$ is similar) follows the process as:
\begin{equation}
\begin{aligned}
|\psi_{+}\rangle \stackrel{U_{1}}{\longrightarrow}|\psi_{+}^{\rm{B}}\rangle \stackrel{U_{2}}{\longrightarrow} |\psi_{+}^{\rm{D}}\rangle \stackrel{U_{3}}{\longrightarrow} e^{i(\frac{\gamma}{2}+\pi)}|\psi_{+}\rangle,
\end{aligned}\label{eq:evpath}
\end{equation}
where
\begin{equation}
\begin{aligned}
|\psi_{+}^{\rm{B}}\rangle&=\cos \frac{\pi}{4}|0\rangle-i\sin \frac{\pi}{4}|1\rangle,\\ |\psi_{+}^{\rm{D}}\rangle&=e^{i \pi} e^{i\frac{\gamma}{2}}\left(\cos \frac{\pi}{4}|0\rangle+i\sin \frac{\pi}{4}|1\rangle\right).
\end{aligned}\label{eq:evpath2}
\end{equation}
Specifically, the dressed state starts from the given point A at the initial time $T_{\rm{A}}$. Under the action of $U_{1}(T_{\rm{B}},T_{\rm{A}})$, it travels along the longitude denoted by A-B to point B at $T_{\rm{B}}$ and the state turns to be $|\psi_{+}^{\rm{B}}\rangle$. Then it evolves to $|\psi_{+}^{\rm{D}}\rangle$ along the path denoted by B-C-D due to $U_{2}(T_{\rm{D}},T_{\rm{B}})$. Here the evolution path B-C-D is not along the longitude but a specific geodesic of the Bloch sphere. Finally, it goes back to the starting point A along path D-E-A owing to $U_{3}(T,T_{\rm{D}})$, where the path shares the same longitude as that of path A-B. The overall effect of $U(T,{T_{\rm{A}}})$ is to drive the dressed state $|\psi_{\pm}\rangle$ to fulfill a cyclic evolution with the path A-B-C-D-E-A and thus obtain a corresponding global phase $\pm(\frac{\gamma}{2}+\pi)$. Therefore, $U_{\rm{g}}(\hat{r},{\gamma})$ can also be described as \begin{equation}
\begin{aligned}
U_{\rm{g}}(\hat{r},{\gamma})=e^{i(\frac{\gamma}{2}+\pi)}|\psi_{+}\rangle\langle\psi_{+} |+e^{i(-\frac{\gamma}{2}-\pi)}|\psi_{-}\rangle\langle\psi_{-}|.
\end{aligned}\label{eq:statedress}
\end{equation}  One can easily verify that for the designed evolution path in this work, the parallel transport condition \cite{erik.15} for the geometric gate is always satisfied for each segment, i.e.,
%\begin{widetext}
	\begin{equation}
	\begin{array}{lll}
	\langle\psi_{\pm}| {U_1^{\dag}(t,T_{\rm{A}})}H_1(t)U_1(t,T_{\rm{A}})|\psi_{\pm}\rangle=0, \\
%\ \ & T_{\rm{A}} \leqslant t\leqslant T_{\rm{B}} \\
	\langle\psi_{\pm}^{\rm{B}}|{U_2^{\dag}(t,T_{\rm{B}})}H_2(t)U_2(t,T_{\rm{B}})|\psi_{\pm}^{\rm{B}}\rangle=0,\\
% &T_{\rm{B}} < t\leqslant T_{\rm{D}} \\
	\langle\psi_{\pm}^{\rm{D}}|{U_3^{\dag}(t,T_{\rm{D}})}H_3(t)U_3(t,T_{\rm{D}})|\psi_{\pm}^{\rm{D}}\rangle=0.\\ %& T_{\rm{D}} < t\leqslant T \\
	\end{array}
	\end{equation}
%\end{widetext}
Therefore the global phase that is obtained is the pure geometric phase, and $U_{\rm{g}}(\hat{r},{\gamma})$ represents the pure geometric gate. We note that this approach here is different from our previous one in Ref.~\cite{Zhang.2020}, where a silicon-based single-dot spin qubit rather than the singlet-triplet qubit is employed. The single-dot spin qubit there is resonantly driven by the microwave field. For the microwave-driven Hamiltonian, one can easily design the time-dependent phase such that the dressed states can evolve always along the longitude, and the dynamical phase is then canceled out. However, this is not applicable for the singlet-triplet qubit system here. For comparison, the path B-C-D during the second part in this work is no longer a longitude but a specific geodesic line determined by the exchange interaction. Meanwhile, as stated above, the small Rabi frequency in Ref.~\cite{Zhang.2020} would also prolong the gate duration and lead to complexity for the control system.

\begin{figure}
	\includegraphics[width=1\columnwidth]{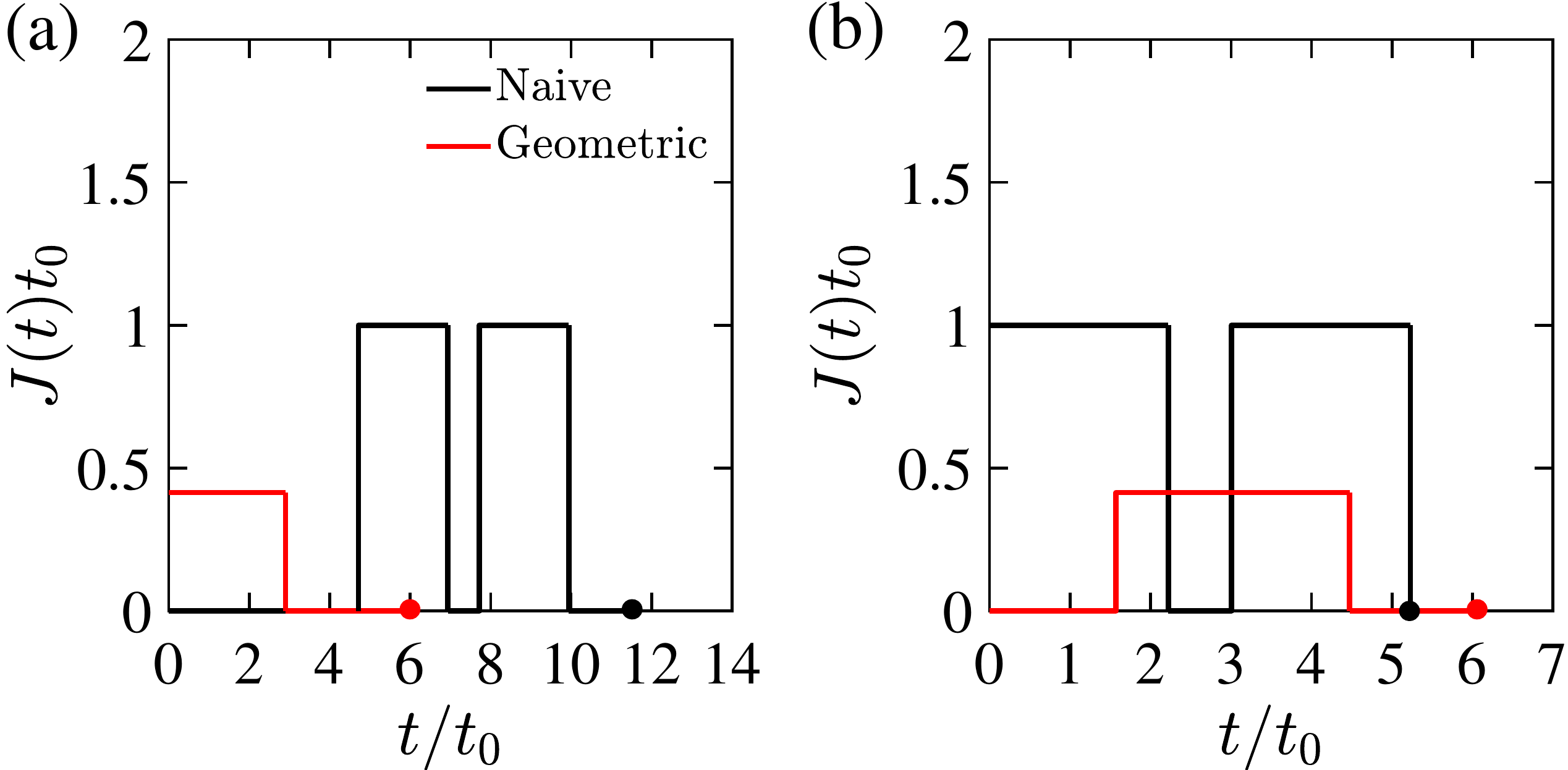}
	\caption{The pulse shapes for the geometric gates $U_{\rm{g}}(\hat{y},\pi/4)$ in (a) and $U_{\rm{g}}(\hat{z},\pi/4)$ in (b) are compared with that of the corresponding naive dynamical gates $U(\hat{y},\pi/4)$  and  $U(\hat{z},\pi/4)$. The black solid line denotes the naive dynamical gate, while the red solid line implies the geometric gate. The time unit is determined by $t_0=1/h$.
	}
	\label{fig:pulse}
\end{figure}

\begin{figure}
	\includegraphics[width=1\columnwidth]{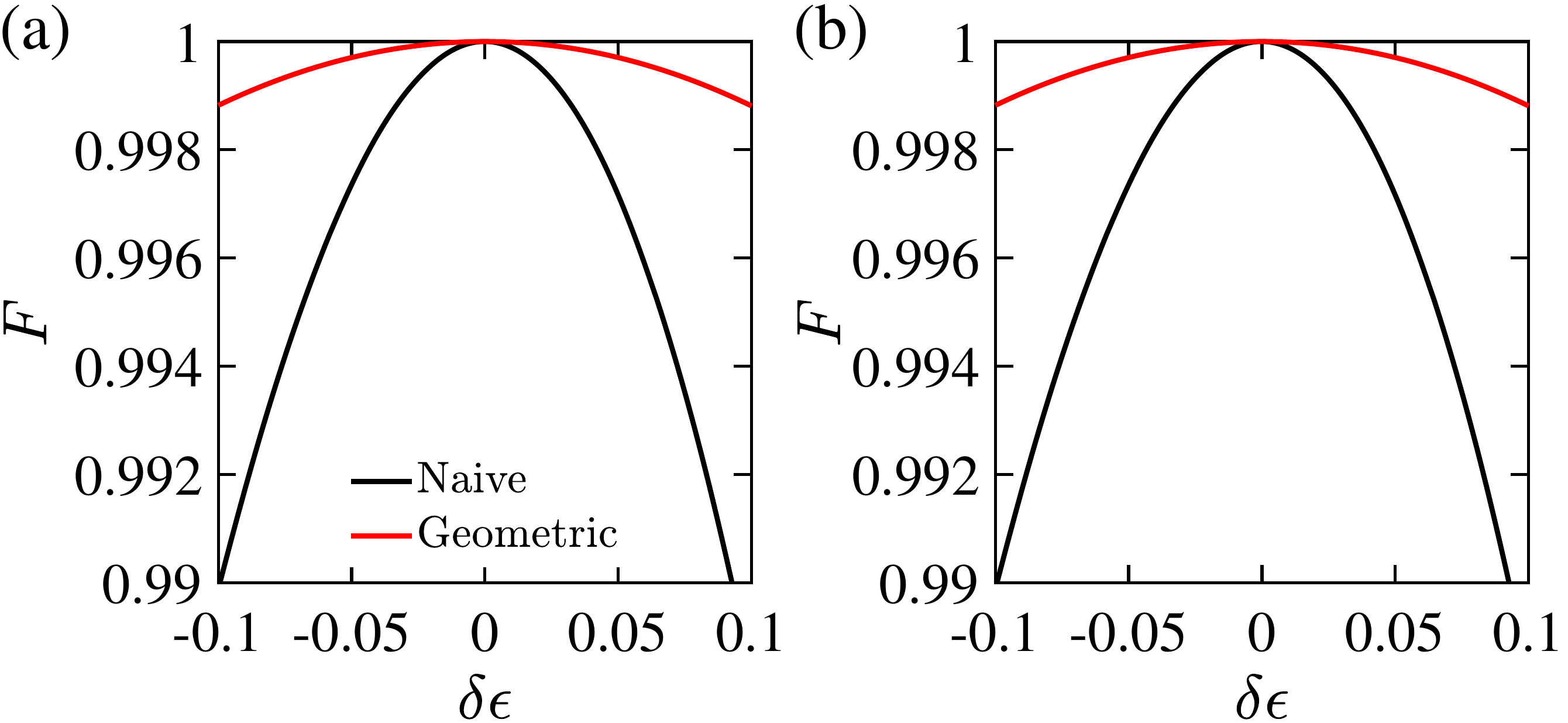}
	\caption{The fidelity is as a function of $\delta\epsilon$, where the charge noise is $\delta J\rightarrow g[J]\delta \epsilon$. The geometric gates $U_{\rm{g}}(\hat{y},\pi/4)$ in (a) and $U_{\rm{g}}(\hat{z},\pi/4)$ in (b) are compared with the naive dynamical gates $U(\hat{y},\pi/4)$  and  $U(\hat{z},\pi/4)$.
	}
	\label{fig:robustness}
\end{figure}

Next we analyze the robustness of the geometric gates compared with the naive dynamical gates. Here we consider two typical rotations, i.e., $U_{\rm{g}}(\hat{y},\pi/4)$ [$U(\hat{y},\pi/4)$] and $U_{\rm{g}}(\hat{z},\pi/4)$ [$U(\hat{z},\pi/4)$]. For the geometric gate $U_{\rm{g}}(\hat{y},\pi/4)$ with $\theta=\pi/2$ and $\gamma=\pi/4$, the used parameters are $J_1=0$, $J_2=\tan(\pi/8)$, and the time duration with respect to each interval is $T_{\rm{AB}}=0$, $T_{\rm{BD}}=\pi\cos{(\pi/8)}$, and $T_{\rm{DA}}=\pi$, while for the corresponding naive dynamical gate $U(\hat{y},\pi/4)$ used in Eq.~(\ref{eq:dynamical naive}) we have $\xi_1=3\pi/2$, $\xi_2=\pi/4$, and $\xi_3=\pi/2$. For $U_{\rm{g}}(\hat{z},\pi/4)$, with $\theta=\pi$, $\gamma=\pi/4$, the parameters used are $J_1=0$, $J_2=\tan(\pi/8)$, $T_{\rm{AB}}=T_{\rm{DA}}=3\pi/2$, and $T_{\rm{BD}}=\pi\cos{(\pi/8)}$. For the naive dynamical gate $U(\hat{z},\pi/4)$ designed using the expression in Eq.~(\ref{eq:rotation-p2}), we have $\xi_0=\pi/4$. The pulse shapes for the geometric and the naive dynamical gates are plotted in Fig.~\ref{fig:pulse}. In Fig.~\ref{fig:robustness} we show their fidelities as a function of  $\delta\epsilon$. Here $\delta\epsilon$ is assumed to be a quasi static noise and its time-dependent effect will be considered later. In all the considered region $-0.1\leqslant\delta\epsilon\leqslant0.1$, which implies $-0.1\leqslant\delta J/J\leqslant0.1$, the geometric gate can outperform its counterpart, i.e., the naive dynamical gate. We notice that for both two naive gates, when $\delta\epsilon$ becomes large the related fidelity drops quickly, whereas the fidelity for the geometric gates varies slowly as $\delta\epsilon$ is increasing. In addition, the advantage of the geometric gates over the naive ones becomes more and more pronounced when $\delta\epsilon$ is large.

\begin{figure}
	\includegraphics[width=0.95\columnwidth]{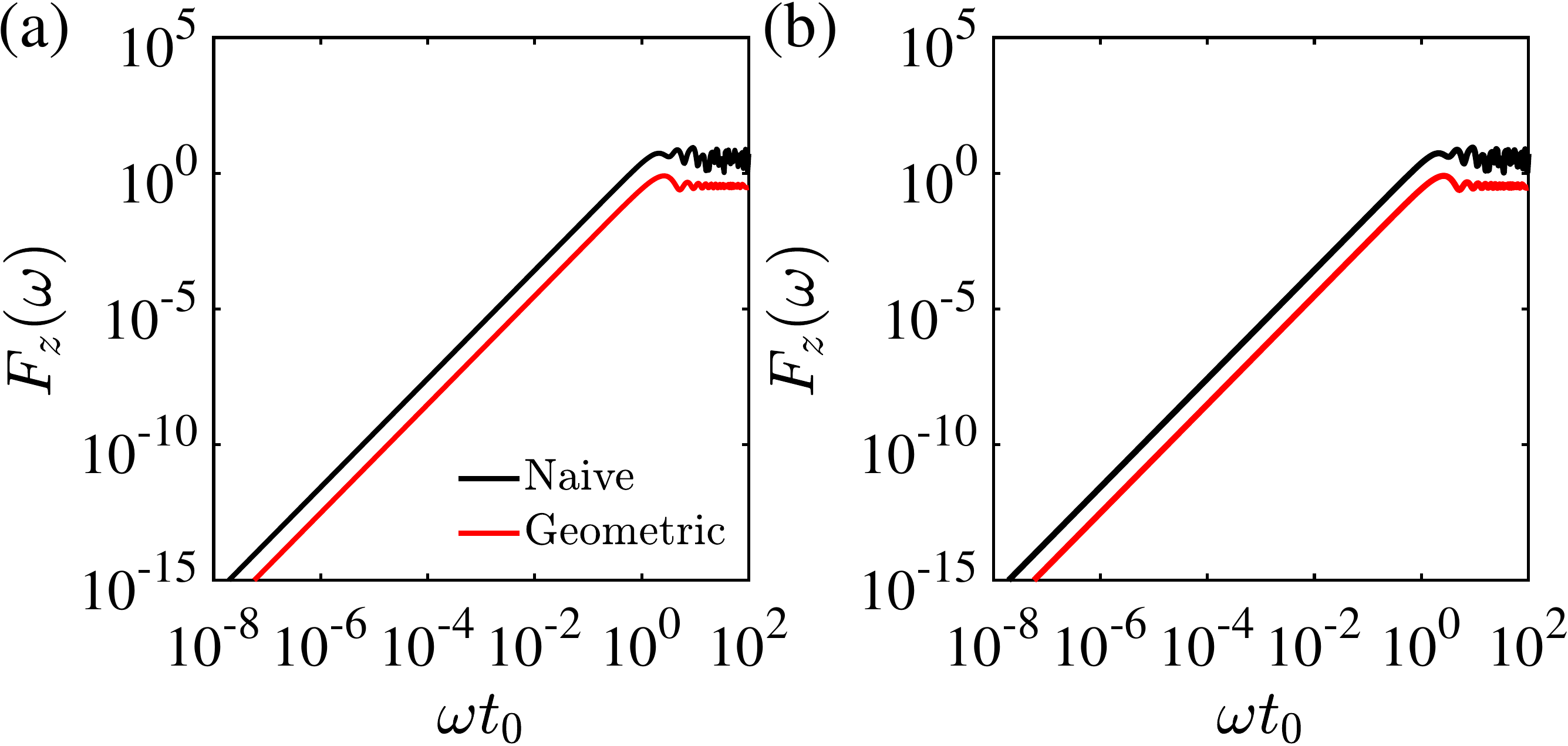}
	\includegraphics[width=0.95\columnwidth]{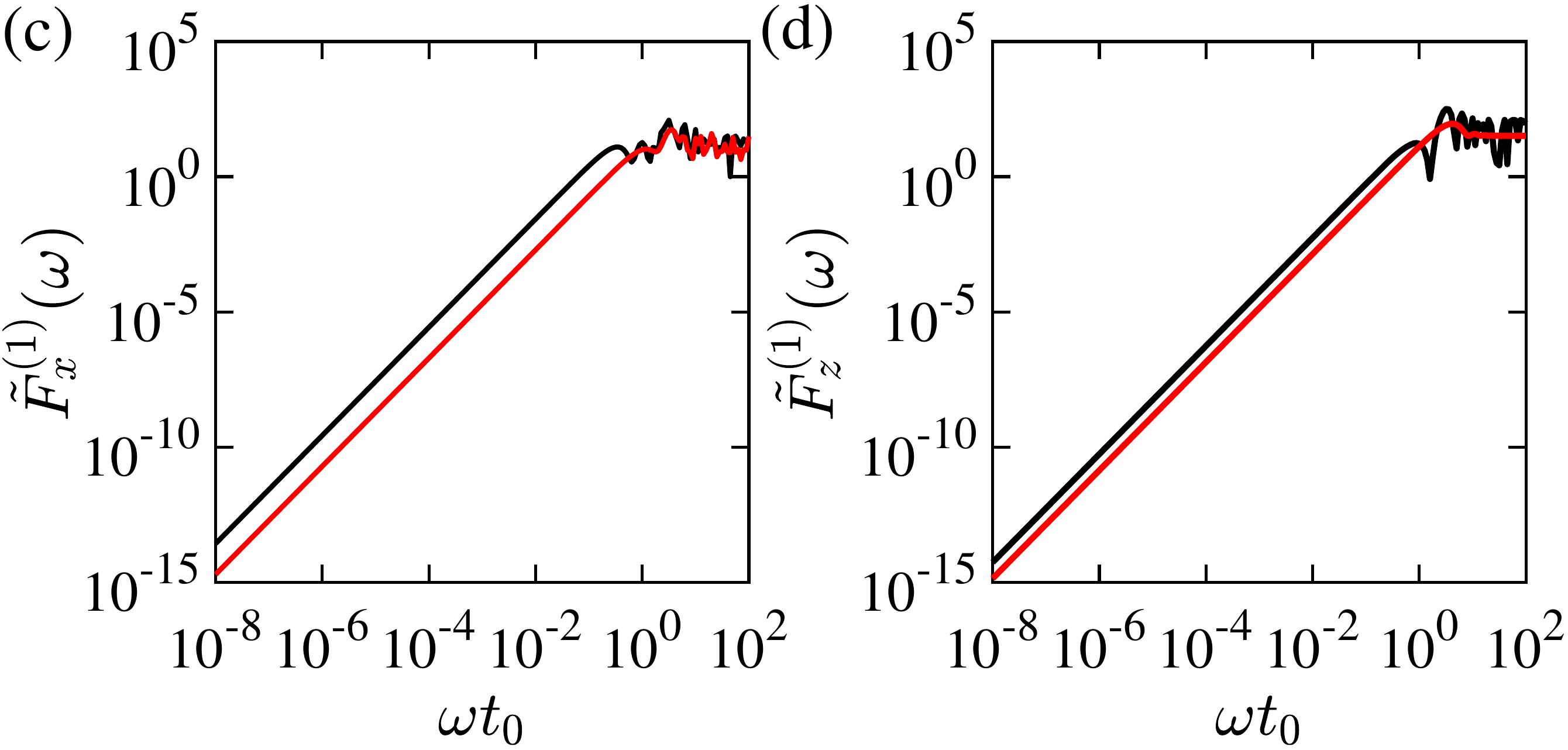}
	\includegraphics[width=0.95\columnwidth]{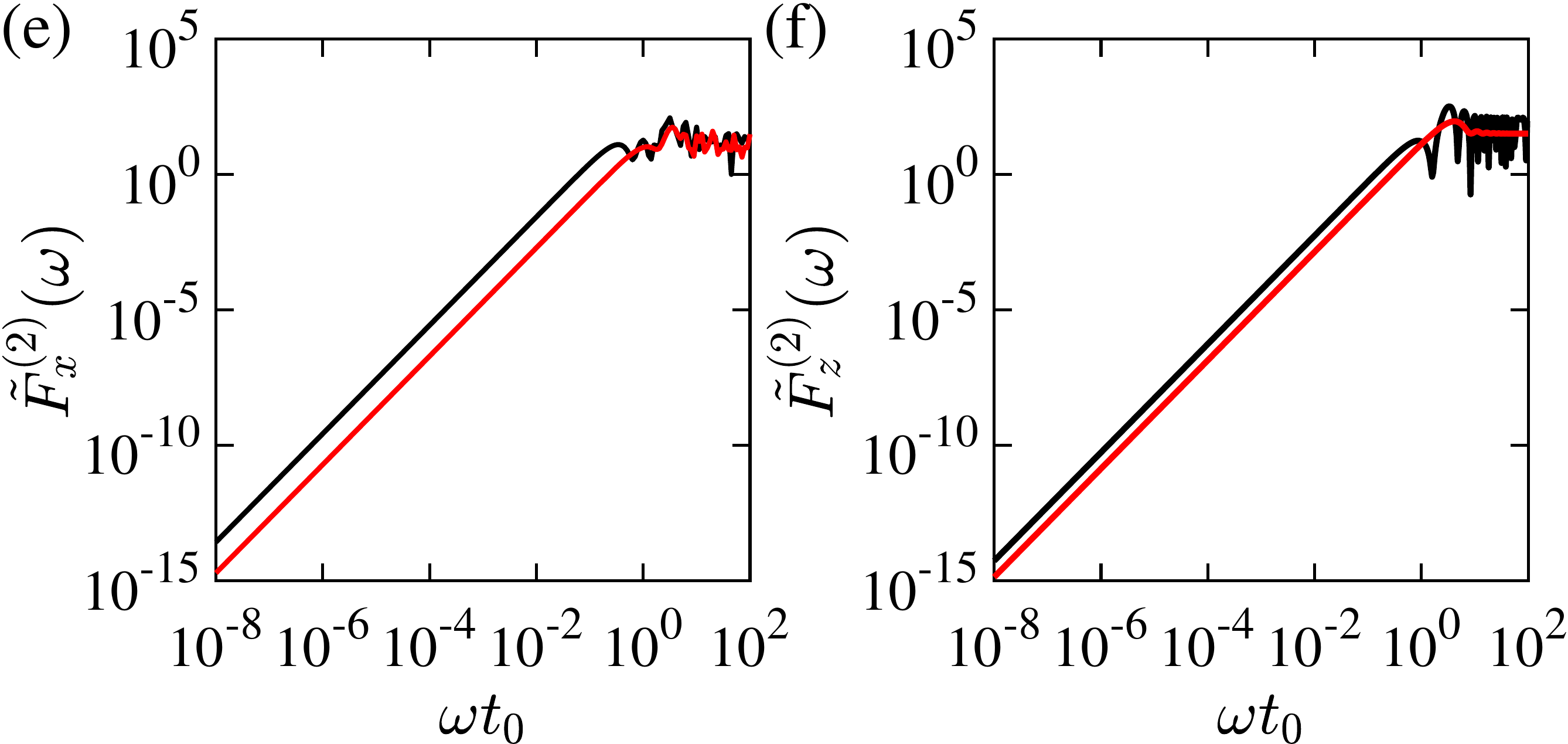}
	\caption{The filter function in panels (a) and (b) are responsible for $U_{\rm{g}}(\hat{y},\pi/4)$ [$U(\hat{y},\pi/4)$] and $U_{\rm{g}}(\hat{z},\pi/4)$ [$U(\hat{z},\pi/4)$], respectively. For the single-qubit case, we only consider the $z$-component noise which is described by $F_{z}(\omega)$. While the filter function for the two-qubit CZ gate is present in (c)- -(f). The superscript $i=1,2$ denotes the $i$th block, and the subscript $i=x,z$ denotes the noise appearing in the effective $\tilde{\sigma}_{i}$ term. The used parameters are $h/(2\pi)=1\  \rm{GHz}$ \cite{Nichol.17},  $S(\omega)=A_{J}/(\omega t_0)^{\alpha}$, with $A_{J} t_{0}=10^{-4}$, and $\alpha=1$. The cutoffs are $\omega_{\rm{ir}}=50\ \rm{kHz}$ and $\omega_{\rm{uv}}=1\  \rm{MHz}$.
	}
	\label{fig:Filter}
\end{figure}

On the other hand, the performance of the geometric gate in the real experimental noise environment in a semiconductor quantum dot remains to be verified, where the noise typically varies over time. For the piecewise control Hamiltonian, the filter function \cite{Green.2012,Green.2013,Paz-Silva.2014} is a powerful tool to evaluate the fidelity for the time-dependent noise. The detail of the filter function is described in Appendix~\ref{appx:filter function}, where it is defined as $F_{i}(\omega)$ ($i=x,z$) in the frequency domain with the noise appearing in the $\sigma_{i}$ term of the Hamiltonian. Note that for the single-qubit case, the charge noise exists in the $\sigma_{z}$ direction. Therefore the filter function  is 
\begin{equation}\label{eq:ffF}
\begin{aligned}
\mathcal{F}=1-\frac{1}{\pi} \int_{\omega_{\mathrm{ir}}}^{\omega_{\mathrm{uv}}} \frac{d\omega}{\omega^{2}} F_{z}(\omega)S(\omega),
\end{aligned}
\end{equation}
where $S(\omega)$ is the noise power spectral density in the frequency domain, and $\omega_{\mathrm{uv}}$ and $\omega_{\mathrm{ir}}$ are the cutoff frequency.
The filter functions for rotation $U_{\rm{g}}(\hat{y},\pi/4)$ [$U(\hat{y},\pi/4)$] and $U_{\rm{g}}(\hat{z},\pi/4)$ [$U(\hat{z},\pi/4)$] are shown in Figs.~\ref{fig:Filter} (a) and \ref{fig:Filter} (b), respectively.
For both two rotations, the lines for the geometric gates are below the ones for the naive dynamical gates, which qualitatively implies the smaller infidelity for the geometric gates compared to their dynamical counterparts.

Here we consider the $1/f^{\alpha}$ noise, which is the typical noise model to describe the time-dependent charge noise in a semiconductor quantum dot. The power spectral density with respect to the charge noise can be written as \cite{yang.2016}
\begin{equation}
S(\omega)=\frac{A_{J}}{(\omega t_0)^{\alpha}},
\label{1fnoise}
\end{equation}
where $A_{J}$ is the noise amplitude, and the exponent $\alpha$ denotes how much the noise is correlated. $t_0=1/h$ is the time unit, and we take $h/(2\pi)=1\ \rm{GHz}$ \cite{Nichol.17}. The noise amplitude can be determined by \cite{Zhang.2017}
\begin{equation}
\begin{aligned}
\int_{\omega_{\mathrm{ir}}}^{\omega_{\mathrm{uv}}} \frac{A_{J}}{\left(\omega t_{0}\right)^{\alpha}}d\omega=\pi\left(\frac{\sigma_{J}}{Jt_0}\right)^{2},
\label{eq:integ}
\end{aligned}
\end{equation}
where $\sigma_{J}$ represents the standard deviation for charge noise. Typically, in a semiconductor quantum dot environment ~\cite{Barnes.2016} we have $\alpha=1$ for the charge noise and the cutoffs are $\omega_{\mathrm{ir}}=50\ \mathrm{kHz}$ and $\omega_{\mathrm{uv}}=1\ \mathrm{MHz}$. In experiments, detuning can be operated via either symmetric (barrier) control or tilt control, which corresponds to $\sigma_{J}/J=0.00426$ for barrier control and $0.0563$ for tilt control, respectively \cite{Martins.2016}. Therefore the noise amplitude region is about $2.0\times10^{-5}\leqslant A_{J}t_{0}\leqslant 3.3\times10^{-3}$. In our simulation we have considered a medium value of $A_{J}t_{0}=10^{-4}$ if not mentioned specifically. We find that the fidelities related to $U_{\rm{g}}(\hat{y},\pi/4)$ and $U(\hat{y},\pi/4)$ are $\mathcal{F}_{\rm{nai}}=99.974\%, \mathcal{F}_{\rm{geo}}=99.997\%$, while for $U_{\rm{g}}(\hat{z},\pi/4)$ and $U(\hat{z},\pi/4)$, the fidelities are also $\mathcal{F}_{\rm{nai}}=99.974\%, \mathcal{F}_{\rm{geo}}=99.997\%$.

\begin{figure}
	\includegraphics[width=1\columnwidth]{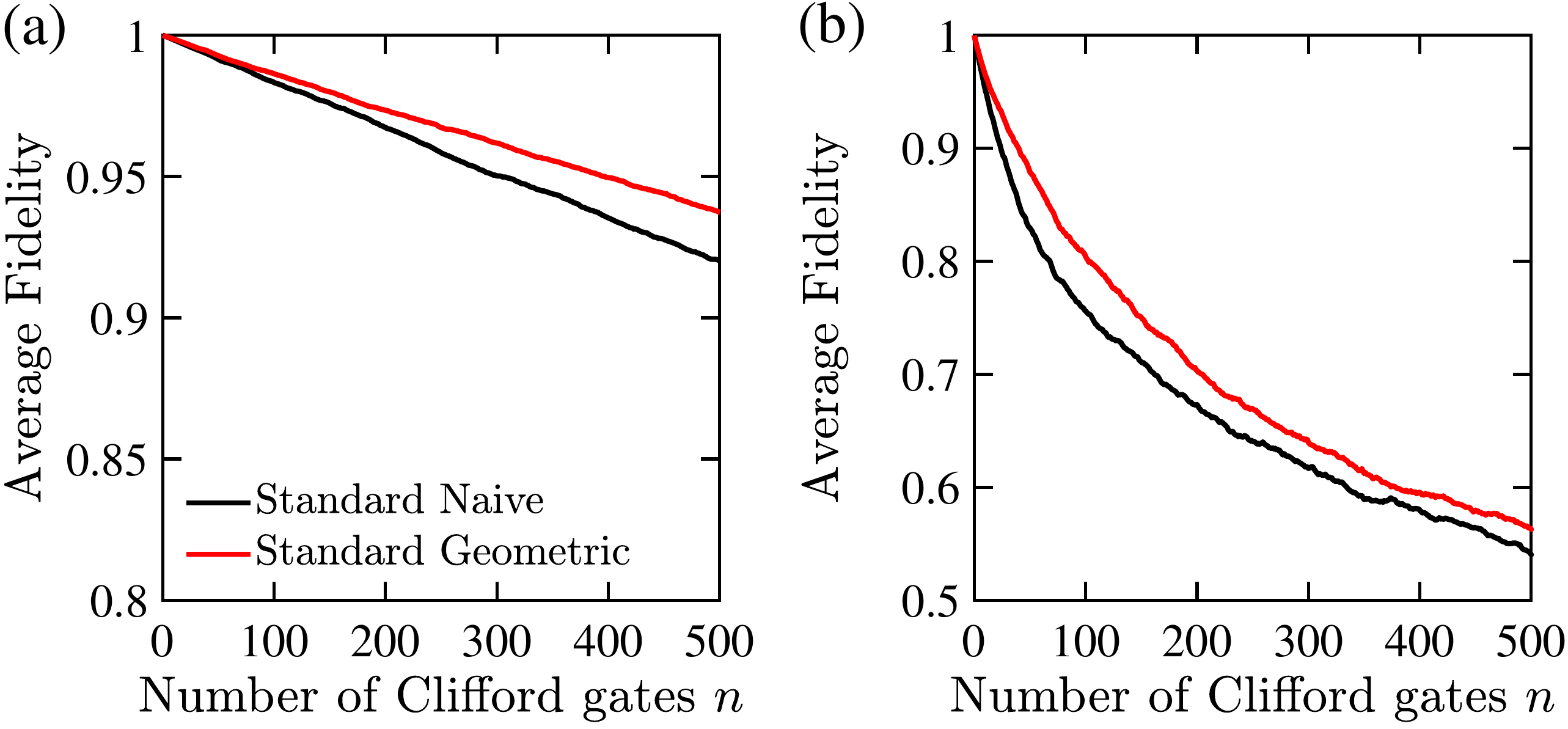}
	\includegraphics[width=1\columnwidth]{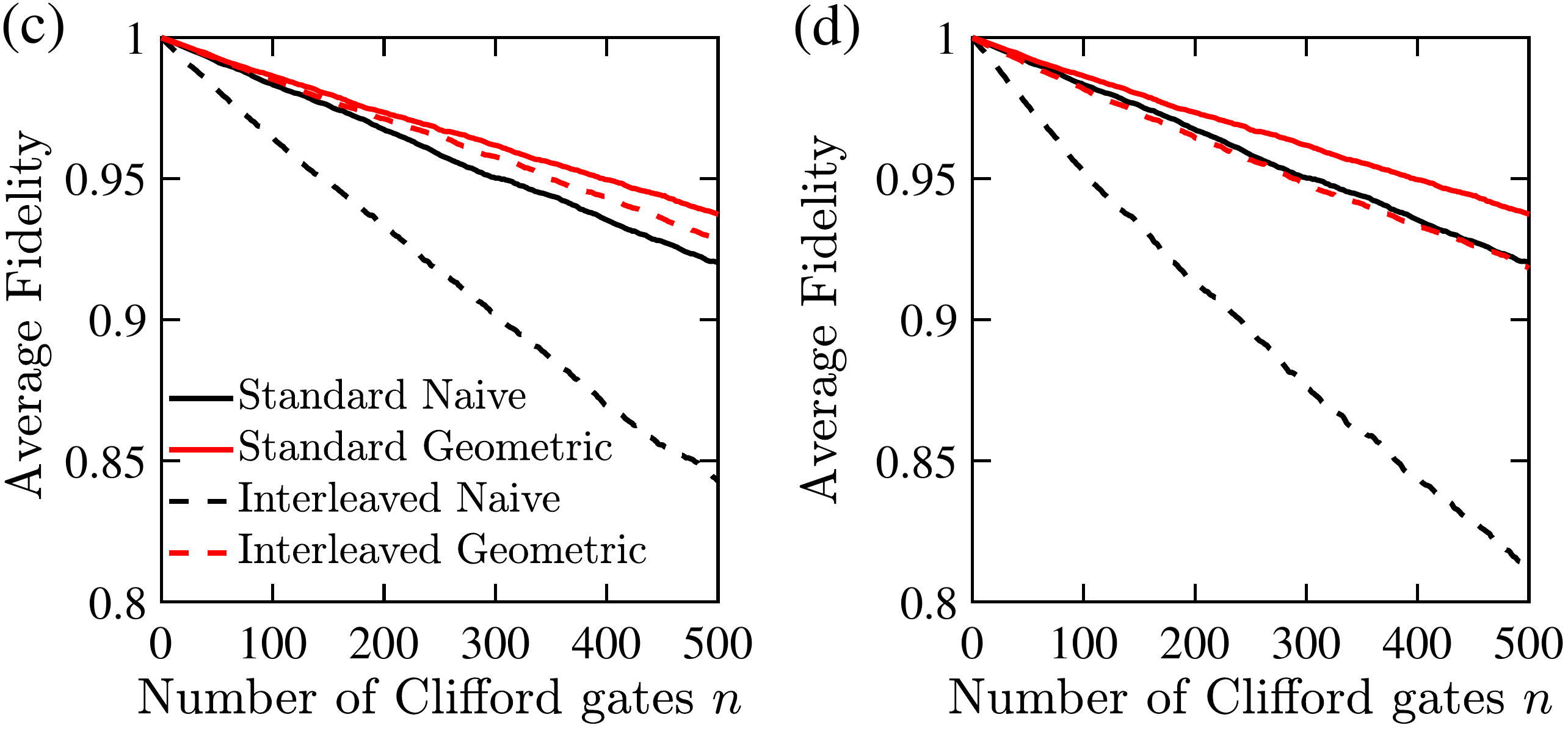}
	\caption{Randomized benchmarking for naive dynamical and geometric gates for $1/f^{\alpha}$ charge noise. The standard randomized benchmarking results are shown in (a) and (b), while the interleaved randomized benchmarking results are shown in (c) and (d) with respect to $U_{\rm{g}}(\hat{y},\pi/4)$ [$U(\hat{y},\pi/4)$] and $U_{\rm{g}}(\hat{z},\pi/4)$ [$U(\hat{z},\pi/4)$], respectively. The used parameters are $h/(2\pi)=1\  \rm{GHz}$ \cite{Nichol.17},  $S(\omega)=A_{J}/(\omega t_0)^{\alpha}$ with $A_{J} t_{0}=10^{-4}$. The noise exponent $\alpha=1$ and 2, corresponding to the left and right column, respectively. The cutoffs are $\omega_{\rm{ir}}=50\ \rm{kHz}$ and $\omega_{\rm{uv}}=1\  \rm{MHz}$.}
	\label{fig:RB}
\end{figure}

Except for the filter function, randomized benchmarking \cite{Emerson.2005,Knill.2008,Easwar.2012} is another effective technique to provide the average error for either all the gates on the Bloch sphere or a specific gate from the Clifford group. The former is related to the standard benchmarking, while the latter the interleaved benchmarking. The basic idea of the standard randomized benchmarking \cite{Wang.2014} is that for a given noise spectrum, we average the fidelity over many gate sequences which are randomly drawn from the single-qubit Clifford group composed of 24 specific gate operations and over random noise realizations. For each run of the sequences in our simulation, the noise is attributed to the $1/f$ form as described in Eq.~(\ref{1fnoise}). The interleaved benchmarking is a slight variant of the standard randomized benchmarking, where the specific gate to be estimated and the randomly chosen Clifford gate sequence interleave with each other \cite{Easwar.2012}. To ensure convergence,   we have averaged the benchmarking over 1000 times of realizations.

The standard randomized benchmarking results are shown in Figs.~\ref{fig:RB} (a) and \ref{fig:RB} (b). By fitting the resulted fidelity curve to $\left(1+e^{-d n}\right) / 2$, one can obtain the average error per gate $d$, where $n$ denotes the number of the used Clifford gates \cite{yang.2016}. The corresponding average fidelity per gate is therefore $\mathcal{F}=1-d$. For the typical value of $\alpha=1$ in Fig.~\ref{fig:RB} (a), the average fidelity for the naive dynamical gate and the geometric gate are $\mathcal{F}_{\rm{nai}}=99.965\%$ and $\mathcal{F}_{\rm{geo}}=99.972\%$. On the other hand, the charge noise spectrum with $\alpha=2$ \cite{Struck.2020} has also been observed in a recent experiment. The corresponding randomized benchmarking result is shown in Fig.~\ref{fig:RB} (b), where the fidelities for the two types of gates are $\mathcal{F}_{\rm{nai}}=99.468\%$ and $\mathcal{F}_{\rm{geo}}=99.565\%$. With the standard randomized benchmarking results in mind, one can further calculate the interleaved randomized benchmarking fidelity \cite{Easwar.2012} as
%\begin{equation}
$\mathcal{F}_{\rm{in}}=1- \left(1-p_{\rm{in}} / p_{\rm{st}}\right)/2$,
%\label{eq:fin}
%\end{equation}
where $p_{\rm{st}}$ and $p_{\rm{in}}$ are the depolarizing parameters for the standard randomized benchmarking and the interleaved randomized benchmarking, respectively, which are determined by $p=e^{-d}$. The interleaved randomized benchmarking results for the gates $U_{\rm{g}}(\hat{y},\pi/4)$ and $U(\hat{y},\pi/4)$ in Fig.~\ref{fig:RB} (c) are $\mathcal{F}_{\rm{nai}}=99.980\%$ and $ \mathcal{F}_{\rm{geo}}=99.998\%$, while for the gates $U_{\rm{g}}(\hat{z},\pi/4)$ and $U(\hat{z},\pi/4)$ in Fig.~\ref{fig:RB} (d) the results are $\mathcal{F}_{\rm{nai}}=99.970\%$ and $ \mathcal{F}_{\rm{geo}}=99.995\%$.

\begin{figure}
	\includegraphics[width=1\columnwidth]{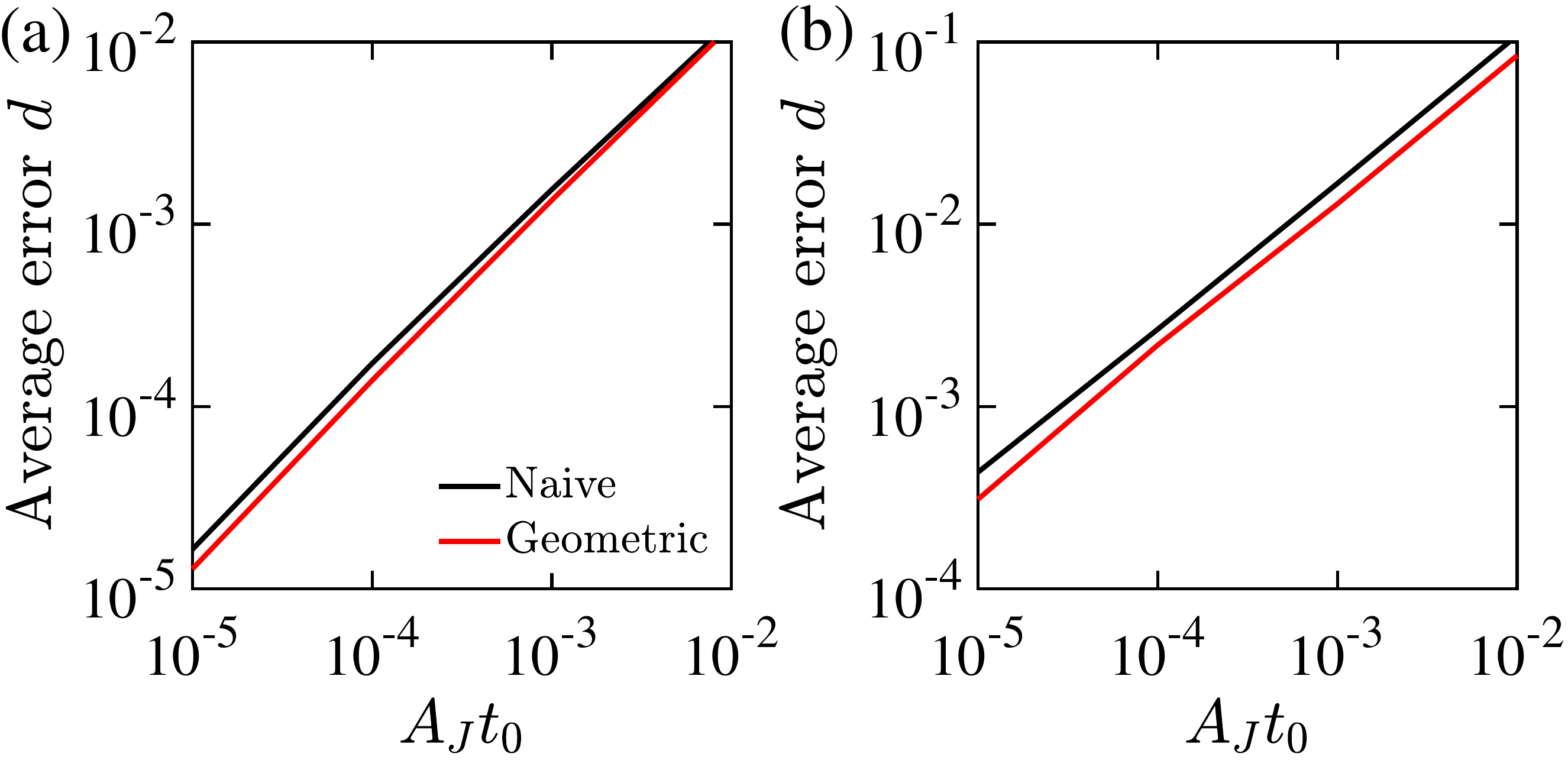}
	\includegraphics[width=1\columnwidth]{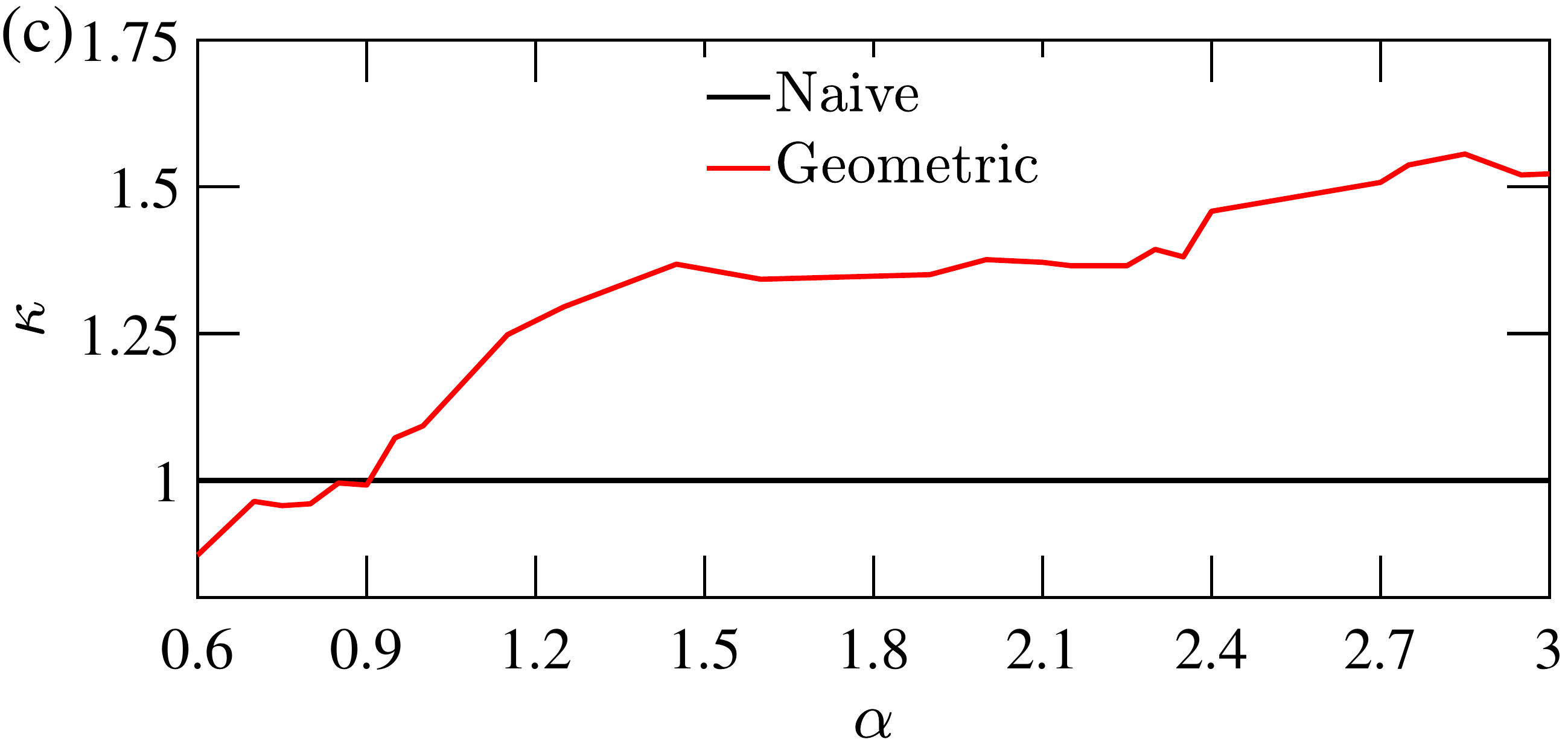}
	\caption{Average error per gate $d$ vs charge noise amplitude $A_{J}t_0$ for $1/f^{\alpha}$ noise, where the noise exponent $\alpha=1$ in (a) and 2 in (b). (c) Improvement ratio $\kappa$ vs $\alpha$.
	}
	\label{fig:error}
\end{figure}

To fully reveal the superiority of the geometric gate, we further consider its performance compared to the naive dynamical one for a wide range of $\alpha$ and noise amplitudes $A_{J}$. In Figs.~\ref{fig:error} (a) and \ref{fig:error} (b) we show the standard average error per gate $d$ as a function of the noise amplitude $A_{J}t_0$ considering the $\alpha$ values related to the recent experiment \cite{Struck.2020}. For $\alpha=1$ in Fig.~\ref{fig:error} (a), when $A_{J}t_0$ is small, the two lines for the naive and geometric gates are almost parallel (the related experimental noise amplitude $A_{J}t_{0}$ is from $10^{-5}$ to $10^{-3}$). However, when $A_{J}t_0$ is large enough (surpassing $10^{-3}$), these two lines are increasingly overlapping, which means there is no improvement for the naive dynamical gates. For $\alpha=2$ in Fig.~\ref{fig:error} (b), the line standing for the geometric gate is lower than the one for the naive dynamical gates in the whole considered noise amplitude. This means the geometric gate is more powerful for large $\alpha$. In Fig.~\ref{fig:error} (c) we further introduce an improvement ratio $\kappa$, which is defined as the error of the naive dynamical gates divided by that of the geometric ones in the small noise amplitude region. We can see that this improvement ratio $\kappa$ is increasing as $\alpha$ becomes larger. When $\alpha\leqslant0.9$, $\kappa$ is less than 1, which means the geometric gate performs worse. For the typical noise exponent $\alpha=1$, we have $\kappa=1.1$, while for the case $\alpha=2.85$, the improvement ratio is 1.56.

\section{Two-qubit geometric gates}\label{sec:twoqubit}

The Hamiltonian of the exchange-coupled two singlet-triplet qubits, as shown in Fig.~\ref{fig:path}(b), is in an Ising-like form \cite{EdwinBarnes.2022}, namely, the qubits are coupled via the form of $\sigma_{z}\otimes\sigma_{z}$ interaction. This model is also applied for the capacitively coupled charge qubits in the semiconductor quantum dot \cite{Shinkai.09}, the superconducting transmon qubits operating in the dispersive regime \cite{Collodo.20}, and also the nuclear magnetic resonance (NMR) system \cite{Vandersypen.05}. For our case, the two-qubit Hamiltonian for the two double-dot systems is \cite{Klinovaja.2012,Li.2012}
\begin{widetext}
\begin{equation}\label{eq:Hamiltonian2}
\begin{aligned}
H_{d0}
&=\frac{J_{12}}{2}\tilde{\sigma}_a^x \otimes \tilde{I}_b +\frac{J_{34}}{2}\tilde{I}_a\otimes\tilde{\sigma}_b^x -\frac{J_{23}}{4}\tilde{\sigma}_a^z\otimes\tilde{\sigma}_b^z+h_a\tilde{\sigma}_a^z\otimes \tilde{I}_b+h_b\tilde{I}_a\otimes\tilde{\sigma}_b^z   \\
&=\left(\begin{array}{cccc}
-J_{23}/4+h_b+h_a        & J_{34}/2   & J_{12}/2             & 0           \\
J_{34}/2   & J_{23}/4-h_b+h_a       & 0            & J_{12}/2           \\
J_{12}/2          & 0           &  J_{23}/4+h_b-h_a         & J_{34}/2    \\
0          & J_{12}/2           & J_{34}/2     & -J_{23}/4-h_b-h_a
\end{array}\right).
\end{aligned}
\end{equation}
For convenience, here we have redefined the basis states as $\{|\tilde{0}\tilde{0}\rangle,\ |\tilde{0}\tilde{1}\rangle,\ |\tilde{1}\tilde{0}\rangle,\ |\tilde{1}\tilde{1}\rangle\}=\{| \uparrow_a\downarrow_a,\uparrow_b\downarrow_b\rangle,\ | \uparrow_a\downarrow_a,\downarrow_b\uparrow_b\rangle,\ | \downarrow_a \uparrow_a,\uparrow_b\downarrow_b\rangle,\ |\downarrow_a\uparrow_a,\downarrow_b\uparrow_b\rangle\}$. In the following we use this basis state to describe the two-qubit operation. (The operation under the logical basis is shown in Appendix~\ref{appx:logic}.) The Pauli matrices are thus slightly different from before:
\begin{equation}
\begin{aligned}
\tilde{\sigma}_i^x &= |\uparrow\downarrow\rangle_i\langle\downarrow\uparrow|_i + |\downarrow\uparrow\rangle_i\langle\uparrow\downarrow|_i, \\
\tilde{\sigma}_i^z &= |\uparrow\downarrow\rangle_i\langle\uparrow\downarrow|_i - |\downarrow\uparrow\rangle_i\langle\downarrow\uparrow|_i, \\
\tilde{I}_i&=|\uparrow\downarrow\rangle_i\langle\uparrow\downarrow|_i + |\downarrow\uparrow\rangle_i\langle\downarrow\uparrow|_i,
\end{aligned}
\end{equation}
where $i=a,b$ denotes the qubit number, and $h_{i}$ is the gradient for each qubit. One can easily find that $\tilde{\sigma}_i^x=\sigma_i^z$ and $\tilde{\sigma}_i^z=\sigma_i^x$. $J_{k,k+1} (k=1,2,3)$ denotes exchange couplings between neighboring dots. Assuming $ h_a,\ h_b\ll J_{23}$ \cite{Li.2012} and setting $J_{12}=0$, $H_{d0}$ turns to be a block-diagonal matrix:
\begin{equation}\label{eq:Hamiltonian3}
\begin{aligned}
H_d=\left(\begin{array}{cccc}
 -J_{23}/4        & J_{34}/2   & 0            & 0           \\
 J_{34}/2   & J_{23}/4       & 0            & 0           \\
  0          & 0           &  J_{23}/4         & J_{34}/2    \\
  0          & 0           & J_{34}/2     & -J_{23}/4
\end{array}\right).
\end{aligned}
\end{equation}
In this way we can decompose $H_{d}$ into two independent subsystems with each subsystem being a $2 \times 2$   matrix:
\begin{equation}
H_{d1}=\left(\begin{array}{ll}
-J_{23}/4 & J_{34}/2  \\
J_{34}/2 & J_{23}/4  \\
\end{array}\right),
H_{d2}=\left(\begin{array}{ll}
J_{23}/4  & J_{34}/2  \\
J_{34}/2  & -J_{23}/4  \\
\end{array}\right).
\label{eq:25}
\end{equation}
Thus we can treat each block as the single-qubit case and design the corresponding geometric operation similar so that in Sec.~\ref{sec:singlegeo}. For the first block $H_{d1}$, we aim to design a single geometric gate. Similarly to the single-qubit case in Eq.~(\ref{eq:paths}) the three-piece evolution needs to satisfy
\begin{equation}\label{eq:path2}
\begin{aligned}
&\int_{T_{\rm{A}}}^{T_{\rm{B}}}\sqrt{(-\frac{J_{23}^{(1)}}{4})^2 +(\frac{J_{34}}{2})^2}dt=\frac{\frac{\pi}{2}-\chi}{2}+2 m_1\pi, & T_{\rm{A}} \leqslant t \leqslant T_{\rm{B}}   \\
&\int_{T_{\rm{B}}}^{T_{\rm{D}}}\sqrt{(-\frac{J_{23}^{(2)}}{4})^2 +(\frac{J_{34}}{2})^2}dt=\pi/2, & T_{\rm{B}} < t \leqslant T_{\rm{D}}  \\
&\int_{T_{\rm{D}}}^{T}\sqrt{(-\frac{J_{23}^{(3)}}{4})^2+(\frac{J_{34}}{2})^2}dt =\frac{\frac{\pi}{2}+\chi}{2}+2 m_2\pi, & T_{\rm{D}} < t \leqslant T,
\end{aligned}
\end{equation}
while for the second block $H_{d2}$, the geometric gate requires
\begin{equation}\label{eq:path3}
	\begin{aligned}
	&\int_{T_{\rm{A}}}^{T_{\rm{B}}}\sqrt{(\frac{J_{23}^{(1)}}{4})^2 +(\frac{J_{34}}{2})^2}dt=\frac{\frac{\pi}{2}-\chi}{2}+2 m_1\pi, & T_{\rm{A}} \leqslant t \leqslant T_{\rm{B}}   \\
	&\int_{T_{\rm{B}}}^{T_{\rm{D}}}\sqrt{(\frac{J_{23}^{(2)}}{4})^2+(\frac{J_{34}}{2})^2}dt=\pi/2, & T_{\rm{B}} < t \leqslant T_{\rm{D}}  \\
	&\int_{T_{\rm{D}}}^{T}\sqrt{(\frac{J_{23}^{(3)}}{4})^2+(\frac{J_{34}}{2})^2}dt =\frac{\frac{\pi}{2}+\chi}{2}+2 m_2\pi, & T_{\rm{D}} < t \leqslant T,
	\end{aligned}
\end{equation}
where $\chi$ is similar to $\theta$, as shown in the single-qubit case, whose parameter is determined by the chosen rotation axis as seen below. In addition, we assume $J_{23}^{(1)}=J_{23}^{(3)}$. $m_i\  (i=1,2)$ depends on the values of $\chi$:
\begin{equation}
m_1=\left\{\begin{array}{ll}
1, & \chi > \pi/2  \\
0, & \chi \leqslant \pi/2  \\
\end{array}\right.\ \ m_2=\left\{\begin{array}{ll}
0, & \chi \geqslant -\pi/2  \\
1, & \chi< -\pi/2  \\
\end{array}\right..
\end{equation}
Here we assume $J_{23}^{(i)}$ ($i=1,2,3$) is  time dependent and the others remain unchanged. By setting $J_{23}^{(1)}=J_{23}^{(3)}=0$ and $J_{23}^{(2)}=2 J_{34}\tan{(\gamma/2)}$, one can acquire a two-qubit perfect entangling gate depending on the chosen value of $\gamma$ as
\begin{small}
\begin{equation}\label{eq:U2}
\begin{aligned}
U_{\rm{ent}}=\left(\begin{array}{cccc}
-\cos{\frac{\gamma}{2}}+i\cos{\chi}\sin{\frac{\gamma}{2}}& -\sin{\frac{\gamma}{2}}\sin{\chi} & 0& 0           \\
\sin{\frac{\gamma}{2}}\sin{\chi}& -\cos{\frac{\gamma}{2}}-i\cos{\chi}\sin{\frac{\gamma}{2}}  & 0& 0         \\
0& 0& -\cos{\frac{\gamma}{2}}-i\cos{\chi}\sin{\frac{\gamma}{2}} & \sin{\frac{\gamma}{2}}\sin{\chi}   \\
0& 0& -\sin{\frac{\gamma}{2}}\sin{\chi}      & -\cos{\frac{\gamma}{2}}+i\cos{\chi}\sin{\frac{\gamma}{2}}
\end{array}\right),
\end{aligned}
\end{equation}
\end{small}
\end{widetext}
where the so-called perfect entangling gate can generate the maximally entangled states, e.g., the $\footnotesize{\text{CNOT}}$ gate \cite{Calderon-Vargas.2015}. Generally, whether a two-qubit gate belongs to a perfect entangling gate can be verified by calculating the local invariants with respect to the matrix of this gate. A detailed description of the local invariant is given in Appendix~\ref{appx:entangling}, where the local invariant $G_{i}$ ($i=1,2,3$) is defined. When taking $\gamma=\pi/2$, we calculate the local invariants of $U_{\rm{ent}}$: $G_1=G_2=0,G_3=1$, which satisfies the condition for the perfect entangling operation \cite{Calderon-Vargas.2015}. It is of great interest that if we further take $\chi=0$, $U_{\rm{ent}}$ is equivalent to a controlled-phase ($\footnotesize{\text{CZ}}$) gate: \cite{Watson.2018}
\begin{equation}\label{eq:UCZ}
\begin{aligned}
U_{\rm{CZ}}=e^{-i\frac{\gamma}{2}}\left(\begin{array}{cccc}
1& 0 & 0& 0           \\
0& e^{i\gamma}  & 0& 0         \\
0& 0& e^{i\gamma} & 0   \\
0& 0& 0      & 1
\end{array}\right).
\end{aligned}
\end{equation}The control Hamiltonian in Eq.~(\ref{eq:Hamiltonian3}) can alternatively implement a dynamical two-qubit gate, where the method is similar to the case in Sec.~\ref{sec:model}. For the first block, arbitrary rotation can be obtained via a composite pulse sequence like Eq.~(\ref{eq:dynamical naive}),
\begin{equation}
\begin{aligned}
U_{d1}(\hat{\tilde{r}},\eta)=& U(\hat{\tilde{x}},\eta_1)U(-\hat{\tilde{z}}+\hat{\tilde{x}},\pi)U(\hat{\tilde{x}},\eta_2) \\
& U(-\hat{\tilde{z}}+\hat{\tilde{x}}, \pi)U(\hat{\tilde{x}},\eta_3),
\label{eq:dynamical naive21}
\end{aligned}
\end{equation}
while for the second block,
\begin{equation}
\begin{aligned}
U_{d2}(\hat{\tilde{r}},\eta)=& U(\hat{\tilde{x}},\eta_1)U(\hat{\tilde{z}}+\hat{\tilde{x}},\pi)U(\hat{\tilde{x}},\eta_2) \\
		& U(\hat{\tilde{z}}+\hat{\tilde{x}}, \pi)U(\hat{\tilde{x}},\eta_3).
\label{eq:dynamical naive22}
\end{aligned}
\end{equation}
To acquire the dynamical perfect entangling gate $U_{\rm{ent}}$, we set $\eta_1=\pi+\chi$, $\eta_2=\gamma$, and $\eta_3=\pi-\chi$.

Normally, it is difficult to perform two-qubit randomized benchmarking simulation, since there are more than 10000 elements in the two-qubit Clifford group. In this way we consider calculating the fidelity of $U_{\rm{ent}}$ via the filter function, as similar for the single-qubit case. The charge noise leads to the error in both the effective $\tilde{\sigma}_{z}$ and $\tilde{\sigma}_{x}$ terms of the Hamiltonian for each block: $\delta J_{23} \propto J_{23}  $ and $\delta J_{34} \propto J_{34} $. For simplicity, here we assume these two noise sources are independent of each other. For each block, the fidelity is
\begin{equation}
\begin{aligned}
\mathcal{F}^{(i)}\simeq& 1-\frac{1}{\pi}  \int_{\omega_{\mathrm{ir}}}^{\omega_{\mathrm{uv}}} \frac{\mathrm{d} \omega}{\omega^{2}} [S(\omega) \tilde{F}_{x}^{(i)}(\omega)+S(\omega) \tilde{F}_{z}^{(i)}(\omega)],
\label{eq:f3}
\end{aligned}
\end{equation}
where $\mathcal{F}^{(i)}$ ($i=1,2$) denotes the fidelity for the $i$th block, while $\tilde{F}_{x}^{(i)}$ and $\tilde{F}_{z}^{(i)}$ represent the $\tilde{x}$- and $\tilde{z}$-component filter function for the corresponding block. The filter function results are shown in Figs.~\ref{fig:Filter} (c)- -\ref{fig:Filter}(f). We find the geometric lines are under those for the naive dynamical gates. This indicates the geometric gates in each block have higher fidelity compared to the naive gates. With the fidelity in each block in mind, we can further calculate the fidelity for the entire evolution matrix. The specific expression is derived in Appendix~\ref{appx:two qubit fidelity}, where
\begin{equation}\label{eq:fidelity5}
\mathcal{F}=\frac{1}{5}+\frac{1}{5}(\mathcal{F}^{(1)}+\mathcal{F}^{(2)})^2.
\end{equation}In Table~\ref{ap:table} we show the fidelity for several values of $\chi$. For all cases, the fidelities related to the geometric gates are surpassing their dynamical counterparts. Considering the region of $-\pi/2 \leqslant \chi \leqslant \pi/2$, the fidelity for the geometric gates can surpass 99\%. Specifically, when taking $\chi=-\pi/2$, the fidelity of the geometric gate is with its largest value of $99.508\%$.

\begin{table}
	\caption{Fidelity of the perfect entangling gate $U_{\rm{ent}}$ for several $\chi$, where $\gamma=\pi/2$. When taking $\chi=0$, $U_{\rm{ent}}$ is equivalent to a $\footnotesize{\text{CZ}}$ gate \cite{Watson.2018}. $\mathcal{F}_{\rm{nai}}^{(i)}$ and $\mathcal{F}_{\rm{geo}}^{(i)}$ imply the fidelity for the naive dynamical and geometric gates for the $i$th block, respectively. $\mathcal{F}_{\rm{nai}}$ and $\mathcal{F}_{\rm{geo}}$ are the overall fidelities.}
	\scalebox{0.85}{
		\begin{tabularx}{9.5cm}{ccccccc}
			\hline
			\hline
			$\chi$ \ \ \ & $\mathcal{F}^{(1)}_{\rm{nai}}$ \ \ \ & $\mathcal{F}^{(1)}_{\rm{geo}}$ \ \ \ & $\mathcal{F}^{(2)}_{\rm{nai}}$  \ \ \ & $\mathcal{F}^{(2)}_{\rm{geo}}$  \ \ \ & $\mathcal{F}_{\rm{nai}}$  \ \ \  & $\mathcal{F}_{\rm{geo}}$  \\
			\hline
			$-\pi/2$ & 98.538\% & 99.692\% & 98.538\% & 99.692\% & 97.678\% & 99.508\%
			\\ 	
			$ -\pi/4$ & 98.525\% & 99.688\% & 98.525\% & 99.688\% & 97.658\% & 99.503\%
			\\
			$0$ & 98.509\% & 99.685\% &  98.509\% & 99.685\% & 97.633\% & 99.496\%
			\\
			$\pi/4$ & 98.490\% & 99.680\% & 98.490\%  & 99.678\% & 97.603\% & 99.488\%
			\\
			$\pi/2$ & 98.451\% & 99.674\% & 96.859\%  & 98.451\% & 97.540\% & 99.479\%
			\\
			\hline
			\hline
	\end{tabularx}}
	\label{ap:table}
\end{table}

\section{Conclusion }\label{sec:conclusion}

In conclusion, we have  proposed a framework to realize nonadiabatic geometric gates for singlet-triplet qubits in semiconductor quantum dots. By only modulating the time-dependent exchange interaction between neighboring quantum dots, both single- and two-qubit geometric gates can be implemented without introducing an extra microwave field. The results clearly shown that the achieved geometric gate is not only superior to its counterpart, namely, the naive dynamical gate, with a high fidelity surpassing 99\%, but can also realize high-speed gate operation with a gating time of nanoseconds. Our result indicates the superiority of the geometric gate, which has great potential to implement robust quantum computing.

\section*{ACKNOWLEDGMENTS}\label{sec:ack}
This work was supported by the National Natural Science Foundation of China (Grant No. 11905065, 11874156), and the Science and Technology Program of Guangzhou (Grant No. 2019050001).

 \appendix

 %\setcounter{equation}{0}

 %\section{Dynamical and geometric phase}\label{appx:phase}

\section{Filter function}\label{appx:filter function}

To calculate the filter function, one needs to use a control matrix $\boldsymbol{R}(t) \equiv\left[\boldsymbol{R}_{x}(t), \boldsymbol{R}_{y}(t), \boldsymbol{R}_{z}(t)\right]^{\mathrm{T}}$ with $\boldsymbol{R}_{j}(t) \equiv\left[R_{j x}(t), R_{j y}(t), R_{j z}(t)\right]$ \cite{Green.2013}, each component of which has the form
\begin{equation}
\begin{aligned}
R_{j k}(t)\equiv[\boldsymbol{R}(t)]_{j k}=\frac{\operatorname{Tr}[U_{c}^{\dagger}(t) \sigma_{j} U_{c}(t) \sigma_{k}]}{2},
\label{eq:A1}
\end{aligned}
\end{equation}
where $j, k \in\{x, y, z\}$. Here the evolution operator $U_{c}(t)$ is the solution to the noise-free Schrodinger equation $i\frac{\partial U_{c}(t)}{\partial t}=H_c(t)U_{c}(t)$, with $H_c$ denoting the noise-free Hamiltonian. For our case, the defined control matrix is slightly different from the original one. This is due to the fact that the charge noise enters the Hamiltonian via $\delta J(t)=g[J(t)] \delta \epsilon(t)$, where the fluctuation $\delta \epsilon(t)$ rather than $J(t)$ exhibits the $1/f$-type noise spectrum. Therefore we consider
\begin{equation}
\begin{aligned}
R_{j k}(t)=g[J(t)]\frac{\operatorname{Tr}[U_{c}^{\dagger}(t) \sigma_{j} U_{c}(t) \sigma_{k}]}{2}
\label{eq:A}.
\end{aligned}
\end{equation}
On the other hand, the Fourier transform of the control matrix in the frequency domain is
\begin{equation}
\begin{aligned}
R_{j k}(\omega)=-i\omega\int_{0}^{T} d t R_{j k}(t) e^{i\omega t}.
\label{eq:A2}
\end{aligned}
\end{equation}
For a given noise power spectral density $S_{ij}(\omega)$, the average fidelity is therefore \cite{Green.2013}
\begin{small}
	\begin{equation}
	\begin{aligned}
	\mathcal{F}_{\mathrm{av}} \simeq& 1-\frac{1}{2 \pi} \sum_{i, j, k=x, y, z} \int_{-\infty}^{\infty} \frac{\mathrm{d} \omega}{\omega^{2}} S_{i j}(\omega) {R}_{j k}(\omega) {R}_{i k}^{*}(\omega).
	\label{eq:A3}
	\end{aligned}
	\end{equation}
\end{small}We then show how to calculate the filter function for the geometric gate, which is designed using the piecewise Hamiltonian [see Eq.~(\ref{eq:Hamiltoniian-paths})]. For the  piecewise Hamiltonian, we define the evolution operator during each interval as
\begin{equation}
U_l(t)=e^{-i H_ l (t-t_{l-1})}.
\label{eq:A4}
\end{equation}
Further, we define the corresponding piecewise control matrix in each interval as
\begin{equation}\label{eq:A5}
R_{jk}^{(l)} (t)=[\boldsymbol{R}^{(l)}(t)]_{jk}=J(t_l)\frac{\operatorname{Tr}[U_l^{\dagger}(t) \sigma_{j} U_l(t) \sigma_{k}]}{2},
\end{equation}
and thus we have
\begin{equation}\label{eq:A6}
R^{(l)}_{jk}(\omega)=[\boldsymbol{R}^{(l)}(\omega)]_{jk}=-i\omega\int_{0}^{t_l-t_{l-1}} d t R_{j k}^{(l)}(t) e^{i\omega t}.
\end{equation}
According to Ref.~\cite{Green.2013}, the whole control matrix in the frequency domain is
\begin{equation}\label{eq:A8}
R_{jk}(\omega)=[\boldsymbol{R}(\omega)]_{jk}=\sum_{l=1}^{n}e^{i \omega t_{l-1}} {R}_{j i}^{l}(\omega) \Lambda_{i k}^{(l-1)},
\end{equation}
where
\begin{equation}\label{eq:A7}
{\Lambda}_{jk}^{(l)}=[\mathbf{\Lambda}^{(l)}]_{jk}=\frac{\operatorname{Tr}[P_l^{\dagger}(t) \sigma_{j} P_l(t) \sigma_{k}]}{2},
\end{equation}
with $P_{l}=U_{l}(t_{l})U_{l-1}(t_{l-1}),...,U_{1}(t_{1})$, and $n$ is the number with respect to the piecewise Hamiltonian. For the single-qubit case, we only consider the charge noise in the $\sigma_{z}$ component, and in this way we have
\begin{equation}
\begin{aligned}
\mathcal{F}_{\mathrm{av}}=&1-\frac{1}{\pi} \int_{0}^{\infty} \frac{d \omega} {{\omega^{2}}} S(\omega) F_{z}(\omega),
\label{eq:A9}
\end{aligned}
\end{equation}
where the $z$-component filter function is
\begin{equation}
\begin{aligned}
F_{z}(\omega)=\sum_{i=x, y, z} R_{z i}(\omega) R_{z i}^{*}(\omega).
\label{eq:A10}
\end{aligned}
\end{equation}

\section{Two-qubit gate under logical basis states}\label{appx:logic}
Considering the two types of basis states satisfying $|\rm{T}\rangle=1/\sqrt{2}(|\tilde{0}\rangle+|\tilde{1}\rangle)$ and $|\rm{S}\rangle=1/\sqrt{2}(|\tilde{0}\rangle-|\tilde{1}\rangle)$, namely, $\tilde{\sigma}_i^x=\sigma_i^z$ and $\tilde{\sigma}_i^z=\sigma_i^x$, the transformation between them is
\begin{equation}\label{eq:U0}
\begin{aligned}
U_{0}=\frac{1}{2} \left(\begin{array}{cccc}
1&  1& 1& 1           \\
1& -1& 1& -1         \\
1&  1&-1& -1   \\
1& -1&-1& 1
\end{array}\right).
\end{aligned}
\end{equation}
Using $U_{\rm{ent}}'=U_0^{\dag}U_{\rm{ent}}U_0$, $U_{\rm{ent}}$ in the logical basis states is
 \begin{widetext}
	\begin{small}
		\begin{equation} \label{eq:Uent2}
		\begin{aligned}
		U_{\rm{ent}}'=\left(\begin{array}{cccc}
		-\cos(\gamma/2)& 0 &0 & i\sin(\gamma/2)e^{-i\chi}    \\
		0  & -\cos(\gamma/2)  &i\sin(\gamma/2)e^{i\chi} &0       \\
		0  & i\sin(\gamma/2)e^{-i\chi} &-\cos(\gamma/2)  &0  \\
		i\sin(\gamma/2)e^{i\chi} & 0   & 0    &-\cos(\gamma/2)  
		\end{array}\right).
		\end{aligned}
		\end{equation}
	\end{small}
\end{widetext}

\section{Local invariant}\label{appx:entangling}

The two-qubit gates can be classified into two types of operations, i.e., the local and nonlocal transformations. The representative of the two-qubit local gate is the $\footnotesize{\text{SWAP}}$ gate, while one of the most typical nonlocal gates is the $\footnotesize{\text{CNOT}}$ gate. The $\footnotesize{\text{CNOT}}$ gate can be used to generate the maximally entangled state such that it belongs to the so-called perfect entangling gate \cite{Calderon-Vargas.2015}. Generally, the condition for a two-qubit gate belonging to the perfect entangling gate is verified by the values of the local invariants, which reads as \cite{Calderon-Vargas.2015}
\begin{equation}
\begin{aligned}
&\sin^{2}\chi\leqslant 4\left |{G} \right | \leqslant1 \\
\rm{and}   \\
&\cos{\chi}(\cos{\chi}-G_3)\geqslant 0.
\end{aligned}
\end{equation}
Here, $G=G_1+iG_2=\left |{G}\right | e^{i\chi}$, and
\begin{equation}
\begin{aligned}
&G_1=\mathrm{Re}[\frac{\operatorname{Tr}^2[m(U)]}{16}],  \\
&G_2=\mathrm{Im}[\frac{\operatorname{Tr}^2[m(U)]}{16}],  \\
&G_3=\frac{\operatorname{Tr}^2[m(U)]-\operatorname{Tr}[m^2(U)]}{4},
\end{aligned}
\end{equation}
are termed as the local invariants. Specifically, the unitary and symmetric matrix $m(U)$ depends on the given operator $U$: $m(U)=(Q^{\dag}UQ)^{T}Q^{\dag}UQ$. While $Q$ is the transformation from the standard basic states $\{|00\rangle,\ |01\rangle,\ |10\rangle,\ |11\rangle\}$ to the Bell states $\{|\Phi_1\rangle=1/\sqrt{2}(|00\rangle+|11\rangle),\ |\Phi_2\rangle=i/\sqrt{2}(|01\rangle+|10\rangle),\ |\Phi_3\rangle=1/\sqrt{2}(|01\rangle-|10\rangle),\ |\Phi_4\rangle=i/\sqrt{2}(|00\rangle-|11\rangle)\}$ \cite{Makhlin.2002}, such that
\begin{equation}\label{eq:transformation operation}
\begin{aligned}
Q=\frac{1}{\sqrt{2}}\left(\begin{array}{cccc}
1   & 0   & 0    & i           \\
0   & i   & 1    & 0           \\
0   & i   &-1    & 0    \\
1   & 0   & 0    & -i
\end{array}\right).
\end{aligned}
\end{equation}

\section{Two-qubit fidelity}\label{appx:two qubit fidelity}
The fidelity of the two-qubit gate can be calculated using the formula \cite{Ghosh.2017}
\begin{equation}\label{eq:fidelity}
\mathcal{F}=\frac{\operatorname{Tr}(UU^{\dag}) +(\lvert{\operatorname{Tr}(U_{\rm{ideal}}^{\dag}U)}\rvert)^2}{d(d+1)}.
\end{equation}
Here $U_{\rm{ideal}}$ denotes the ideal target gate operation, $U$ is the actual unitary operation, and $d$ represents the dimension of the Hilbert space. Since the two-qubit Hamiltonian in Eq. (\ref{eq:Hamiltonian3}) is block diagonalized, the corresponding operator $U_{\rm{ideal}}$ can be expressed as
\begin{equation}\label{eq:evo1}
\begin{aligned}
U_{\rm{ideal}}^{\dag}=\left(\begin{array}{cccc}
a_{11}   & a_{12}    & 0            & 0         \\
a_{21}   & a_{22}    & 0            & 0         \\
0        & 0         & a_{33}       & a_{34}    \\
0        & 0         & a_{43}       & a_{44}
\end{array}\right),
\end{aligned}
\end{equation}
and $U$ can be written as
\begin{equation}\label{eq:evo2}
\begin{aligned}
U=\left(\begin{array}{cccc}
b_{11}   & b_{12}    & 0            & 0         \\
b_{21}   & b_{22}    & 0            & 0         \\
0        & 0         & b_{33}       & b_{34}    \\
0        & 0         & b_{43}       & b_{44}
\end{array}\right).
\end{aligned}
\end{equation}
We further define
\begin{equation}
\begin{aligned}
U_{\rm{i}1}^{\dag}=\left(\begin{array}{cc}
a_{11}   & a_{12} \\
a_{21}   & a_{22}
\end{array}\right),\ \  U_{\rm{i}2}^{\dag}=\left(\begin{array}{cc}
a_{33} & a_{34}    \\
a_{43} & a_{44}
\end{array}\right),
\end{aligned}
\end{equation}
and
\begin{equation}
\begin{aligned}
U_{1}=\left(\begin{array}{cc}
b_{11}   & b_{12} \\
b_{21}   & b_{22}
\end{array}\right),\ \   U_{2}=\left(\begin{array}{cc}
b_{33} & b_{34}    \\
b_{43} & b_{44}
\end{array}\right).
\end{aligned}
\end{equation}
Further, we calculate the term of $\operatorname{Tr}(U_{\rm{ideal}}^{\dag}U)$:
\begin{equation}\label{eq:evo3}
\begin{aligned}
\operatorname{Tr}(U_{\rm{ideal}}^{\dag}U)=a_{11}b_{11}+a_{12}b_{21}+a_{21}b_{12}+a_{22}b_{22}  \\
+a_{33}b_{33}+a_{34}b_{43}+a_{43}b_{34}+a_{43}b_{44}.
\end{aligned}
\end{equation}
On the other hand, because the matrix for the operator is diagonalized, each block can be treated as a pseudo-single-qubit gate. In this way, the fidelity with respect to each block can also be calculated as
\begin{eqnarray}\label{eq:fi1}
\mathcal{F}^{(1)}&=& \frac{\operatorname{Tr}(U_{\rm{i}1}^{\dag}U_1)}{2}\notag\\
   &=&\frac{a_{11}b_{11}+a_{12}b_{21}+a_{21}b_{12}+a_{22}b_{22}}{2}   \notag,\\
\mathcal{F}^{(2)}&=& \frac{\operatorname{Tr}(U_{\rm{i}2}^{\dag}U_2)}{2}\notag\\
   &=&\frac{a_{33}b_{33}+a_{34}b_{43}+a_{43}b_{34}+a_{43}b_{44}}{2}.  \notag\\
\end{eqnarray}
By inserting Eqs. (\ref{eq:fi1}) and (\ref{eq:evo3}) into Eq. (\ref{eq:fidelity}), we can get
\begin{small}
\begin{equation}\label{eq:ffF2}
\mathcal{F}
=\frac{4+(\lvert{\operatorname{Tr}(U_{\rm{i}1}^{\dag}U_1) +\operatorname{Tr}(U_{\rm{i}2}^{\dag}U_2)}\rvert)^2}{4\times 5}   \\
=\frac{1}{5}+\frac{(\mathcal{F}^{(1)}+\mathcal{F}^{(2)})^2}{5},
\end{equation}
\end{small}where, we have used $UU^{\dag}=\hat{I}$, while $\hat{I}$ is the identity operator and $\operatorname{Tr}(UU^{\dag})=4$ with $d=4$.

%\bibliography{referenceG}


\end{document}